\documentclass[a4paper,singlecoulomb,12pt]{article}
\usepackage{epsfig,amsmath,amssymb}
\usepackage[colorlinks=true,linktocpage=true,linkcolor=blue,citecolor=blue]{hyperref}
\usepackage{cite}
\usepackage{appendix}
\usepackage{graphicx}
\usepackage{dcolumn}
\usepackage{bm}
\usepackage{subfig}
\usepackage{float}
\usepackage{tabularx}
\usepackage{hhline}
\usepackage{color}
\usepackage{latexsym}
\usepackage{slashed}
\newcommand{\ba}{\begin{eqnarray}}
\newcommand{\ea}{\end{eqnarray}}

\usepackage{caption}

\begin{document}
\title {\large Seebeck and Nernst coefficients of a magnetized 
hot QCD medium with number conserving kernel }

\author{Salman Ahamad Khan\footnote{skhan@ph.iitr.ac.in} and  Binoy Krishna
Patra\footnote{binoy@ph.iitr.ac.in} \vspace{0.3in} \\
Department of Physics, \\
Indian Institute of Technology Roorkee, Roorkee 247667, India}
\date{}
\maketitle
\vskip 0.01in

\begin{flushright}
{\normalsize
%June 21 2008}
}
\end{flushright}
%\vskip 0.1in

\begin{abstract} 
We  study  the thermoelectric 
response of a hot and magnetized QCD medium created in the 
noncentral events at heavy-ion collider experiments. 
The collisional aspects of the 
medium have been 
embedded in the relativistic Boltzmann transport equation
(RBTE) using
Bhatnagar-Gross-Krook (BGK)
 collision integral, which insures the 
particle number conservation, unlike 
the commonly used relaxation time approximation (RTA). We have 
incorporated the thermal medium effects in the guise of 
a quasiparticle model,
where the interaction among the quarks and gluons
is assimilated in the medium dependent masses
of the quarks, which have been evaluated
using imaginary-time formalism of thermal QCD 
with a background magnetic field. In the absence of $B$, 
the Seebeck 
coefficient for individual quark flavors gets slightly
reduced in the BGK term  in comparison to naive RTA,
 while it gets enhanced
for the composite partonic medium. 
In the  strong magnetic 
field ($B$), the BGK term enhances the Seebeck coefficient
for the individual flavors as well as that for medium.
The medium Seebeck coefficient 
rises with the strength of 
quark chemical potential ($\mu$) 
in the absence
as well as that in the strong $B$.   
We observe chirality dependence in the 
 transport coefficients in the weak $B$ as the
 masses of chiral modes become nondegenerate.
 In the case of the $L$ modes, the BGK collision term
 causes slight reduction in the Seebeck coefficient,
 while for $R$ modes both the collision 
 integral produces same results. Nernst coefficient
 gets reduced (enhanced) for $L$ ($R$) chiral
 modes in the BGK term.
\end{abstract}

\section{Introduction}
 A transition from the hadronic matter to 
a deconfined phase of  quark gluon plasma (QGP) 
takes place
 in  heavy-ion 
collision experiments at Relativistic Heavy Ion Collider (RHIC) 
and Large Hadron Collider (LHC). In non-central collisions,  
 a magnetic field (around $m_\pi^2$
at RHIC
\cite{Kharzeev:NPA803'2008} and  $15m_\pi^2$ at 
LHC \cite{Skokov:IJMPA24'2009}) is also 
produced, which persists in the medium for 
a considerable amount of time due to the 
finite electrical conductivity of the medium. This magnetic 
field leads to the modification in the thermodynamical
~\cite{Rath:JHEP1712'2017,Karmakar:PRD99'2019}
and transport properties~\cite{Rath:PRD102'2020,
Kurian:EPJC79'2019,Li:PRD97'2018,Hattori:PRD96'2017,
Chen:PRD101'2020,Nam:PRD87'2013,
Pushpa:PRD105'2022} of the hot and dense quark matter and 
also induces novel phenomena  
 {\em such as} the chiral magnetic 
effect\cite{Fukushima:PRD78'2008,Kharzeev:NPA803'2008},
magnetic and inverse magnetic catalysis\cite{Gusynin:PRL73'1994,Leung:PRD55'1997,
Gusynin:PRD56'1997,Shovkovy:LNP'871'2013}, 
axial magnetic effect\cite{Braguta:PRD89'2014,
Chernodub:PRB'2014},
chiral vortical effect in rotating QGP
\cite{Kharzeev:PRL106'2011, Kharzeev:PPNP88'2016},
the conformal anomaly
 and production of soft photons\cite{Fayazbakhsh:PRD88'2013,
 Basar:PRL109'2012}.
 In addition to this, the dilepton production rate\cite
{Tuchin:PRC88'2013,Bandyopadhyay:PRD94'2016,Sadooghi:ANP376'2017}, 
 dispersion relations
 \cite{Sadooghi:PRD92'2015}, refractive indices 
and decay constants\cite{Fayazbakhsh:PRD86'2012,
Fayazbakhsh:PRD88'2013} have been  explored in the 
magnetic field background.\par

Transport coefficients are crucial input parameters
needed in the dissipative hydrodynamics and transport 
simulation to describe the evolution of the partonic medium
 created post-collision at
RHIC and LHC.  Shear viscosity quantifies the 
response of the 
medium to the transverse momentum gradients while
 bulk viscosity
to the pressure
gradients. Both shear and bulk viscosities  have been 
studied in the magnetic field extensively
in different models~\cite{Rath:PRD102'2020,
Li:PRD97'2018,Hattori:PRD96'2017,
Chen:PRD101'2020,Nam:PRD87'2013}. 
Electrical and thermal conductivities 
 measure  the response of the system to the 
electromagnetic fields and 
thermal gradients in the medium,
respectively. Electrical conductivity 
plays an important role 
in the elongation of the lifetime 
of the magnetic field created in non-central 
collisions, while thermal conductivity
controls the attenuation of sound 
through the Prandtl number. Both  electrical and 
thermal conductivities have been extensively 
studied in phenomenological models as well as 
using perturbative methods~\cite{Kurian:EPJC79'2019,
PV:PRL105'2010,Seung:PRD86'2012,Kharzeev:PPNP75'2014,
Satow:PRD90'2014,Pu:PRD91'2015,Hattori:PRD94'2016}. 
On the other hand, the transport 
coefficient corresponding to the thermoelectric
response is known as Seebeck coefficient, which measures
the ability of any material to convert the 
thermal gradient into the electric current. 
Thermoelectric properties of the materials have 
been mainly studied in the context of
condensed matter physics over the years. There have been 
numerous studies regarding the thermoelectric 
properties of the various condensed matter systems {\em such as}
superconductors~\cite{Pao:9505002,Matusiak:PRB97'2018,
Cyr:PRX7'2017,Gaudart:PSS2'2008}, the graphene
superconductor junction~\cite{Wysokinski:JAP113"2013}, 
 correlated quantum dots coupled to
superconducting electrode~\cite{Wojcik:PRB89'2014}, 
high temperature cuprates~\cite{Seo:PRB90'2014},
 ferromagnet-superconductor hybrid junction~
 \cite{Dutta:PRB96'2017} and low dimensional correlated organic
metals~\cite{Shahbazi:PRB94'2016}.\par

We have recently explored the charge, heat, and momentum
transport coefficients~\cite{Khan:PRD104'2021,Khan:PRD106'2022}.  
Motivated by our earlier studies, we are now interested in 
thermoelectric effects in the strongly 
interacting matter produced 
in the heavy-ion collisions where a thermal gradient is 
present between the central and peripheral regions of the 
fireball. In addition to the temperature gradient,
a finite baryon chemical potential is also needed to 
observe the thermoelectric effect in strongly 
interacting matter unlike the condensed matter systems,
where  only one type of the  charge 
carriers participate in the transport process. Contrary to that, 
in strongly interacting medium, 
both positive and negative charge carriers take part in 
the transport phenomena. In the absence of the 
quark chemical potential, both particles and anti-particles
have equal numbers, so no net thermoelectric effect
is observed. 
The Seebeck effect in the absence of 
a magnetic field has been studied recently for a hot hadron gas 
in hadron resonance gas model~\cite{Bhatt:PRD99'2019} 
and for the QGP phase in the ambit of    
Nambu–Jona Lasinio (NJL) model~\cite{Aman:EPJC82'2022}. In
the  magnetic field background, the thermoelectric
response of the hot QCD medium has been explored 
earlier in~\cite{Das:PRD102'2020,Kurian:PRD103,Zhang:EPJC81'2021,
Dey:PRD102'2020,Dey:PRD104'2021,
Dey:arXiv'2204.06195}. 
 In the earlier works,
authors have used the relativistic kinetic theory approach,
where the collisional effects of the medium
have been incorporated with RTA. 
But, the widely used RTA collision integral 
has a drawback that it violates the conservation 
of the particle number and current. 
Taking this fact into consideration,
 we have used a more realistic
 BGK-type collision integral, which
insures the particle number and current conservation
in the medium. 
 The BGK collision
integral has been used earlier to 
study dielectric functions, dispersion relations
and damping rates of longitudinal and transverse 
modes of a photon in the electromagnetic 
plasma~\cite{Carrington:CJP82'2004}. The authors 
 noticed a small shift in the dispersion relations
towards the lower energies for the collisional case
in comparison to collisionless case. Schenke et 
al.~\cite{Schenke:PRD73'2006} have studied the 
effects of the collisions using the BGK kernel on the 
collective modes of a gluon in the anisotropic 
 QCD medium 
and have observed that incorporation of BGK collision integral 
slows down the growth of the unstable modes. 
The gluonic  collective modes have been also studied 
in anisotropic medium within
 the effective fugacity model~\cite{Kumar:PRD97'2018}
 and suppression of the instabilities were reported
 there also.
 The effects of the collision has been investigated
using BGK term  on the square of the refractive index ($n^2$) 
  and Depine-Lakhtakia index ($n_{DL}$) for the 
   QGP medium 
in ref.~\cite{Jiang:PRD94'2016}. It was noticed that 
 the real and imaginary
parts of the $n^2$ gets changed drametically
compared to collisionless case.
For a small collision
rate, $n_{DL}$ becomes negative in 
certain frequency range and as the 
collision rate increases, the frequency range for $n_{DL}<0$
becomes narrower. The 
wakes phenomenon   
 has been explored for both isotropic~\cite{Chakraborty:JPG34'2007} as well
as anisotropic medium~\cite{Mandal:PRD88'2013}
 and it is observed that the wake structure becomes
 less pronounced in both the cases in comparison 
 to the collisionless plasma. The effect of collisions 
 on the 
  heavy quark energy 
 loss has been investigated  via BGK kernel and it is 
 found that for the same momentum and collision frequency,  
 energy loss gets increased in the BGK case in comparison to the collisionless case for both charm and bottom 
 quarks and further increases as the collision rate is increased~\cite{Cheng:EPJA53'2017}. Authors
 in~\cite{Yousuf:EPJC79'2019} perform similar study 
using the effective fugasity model and considering
  both RTA as well as BGK 
 collision terms. They 
 observed that the energy loss gets reduced 
 in the BGK case as compared to RTA. 
 In addition to 
 these works, the response of stationary and homogenous 
 quark gluon plasma to the background electromagnetic 
  field has been  studied in~\cite{Grayson:PRD106'2022}.
  It was found that the latetime magnetic field
  is mainly determined by the static elecrical 
  conductivity of the medium. A similar kind of study 
was made for the electron positron plasma with time and space dependent magnetic fields~\cite{Formanek:Annphys434'2021}. The 
electric charge transport in a weakly magnetized hot 
QCD medium in the presence of 
 time varying electric field has been investigated in~\cite{Gowthama:PRD103'2021}. 
Both Ohmic and Hall conductivies get enhanced in the 
BGK term as compared to RTA. Similar observations were
noticed in the strong magnetic 
field, where longitudinal electrical 
conductivity becomes larger in the BGK term~\cite{Khan:PRD104'2021}. The momentum  
transport coefficients have been studied in strong
magnetic field with BGK kernel by us in ref.~
\cite{Khan:PRD106'2022} and we notice 
that the shear viscosity gets
enhanced while bulk viscosity is reduced slightly
in comparison to RTA.  
  \par
 Motivated by the earlier works, our main 
objective here is to 
investigate how current conserving BGK collision integral 
modifies  the thermoelectric
transport coefficients {\em namely} Seebeck and 
Nernst coefficients of the hot QCD medium.
 We include the  medium effects
in the framework of a quasi-particle model~\cite{Bannur:JHEP09'2007}, where
medium dependence enters through 
the dispersion relations of the 
quark and gluon quasi-particles.
Quasi-particle models 
are widely used to study the thermodynamical and
transport  properties of the hot QCD medium.  
The masses of the 
quarks have been computed from the pole of the
 propagator resummed using the Dyson-Schwinger equation.
We have employed the perturbative thermal QCD 
in magnetic field background
to calculate the self-energy of the 
quark. We compare
the BGK results with those obtained using 
RTA. We
have explored  two regimes of the magnetic 
field, the strong 
($|q_iB|>>T^2>>m_i^2$) and the
weak magnetic field regime ($T^2>>|q_iB|>>m_i^2$). 
In the magnetic field, 
the motion of the quarks is quantized in 
the transverse direction leading 
to the discrete energy spectrum in 
terms of the Landau levels. When the strength of the 
magnetic field is large, the energy separation between the 
 consecutive Landau levels become 
 large (of the order $\sqrt{|qB|} $),
 consequently, the quarks get confined in the lowest 
 Landau level (LLL) only. Moreover, in the weak 
 magnetic field case, the magnetic field dependence
 enters through the cyclotron frequency. 
% The degeneracy in the masses of the different chiral modes  
% is lifted, as a result, the mass depends 
% on the handedness of the quasiparticles, which leads 
%to the chirality dependence in the 
% transport coefficients~\cite{Pushpa:PRD105'2022,
% Dey:arXiv'2204.06195}.
We found that in the absence 
of the magnetic field, the 
magnitude of the Seebeck coefficient 
for the individual $u$, $d$ and $s$
quarks gets reduced in the BGK collision integral
 while for the composite medium, it 
 gets enhanced. 
 In the strong $B$,  it gets enhanced in BGK
 collision term  for 
 individual flavors as well as for medium.
 In case of the weak magnetic field,
 the Seebeck coefficient is not much sensitive
 to the collision integral and 
 found to be almost similar in both 
 collision terms. In addition,  
  a hall type  Nernst coefficient  
 also appears, which quantifies  the thermoelectric 
 response in the transverse direction. Nernst coefficient
 gets reduced (enhanced) in the BGK term in the case of 
 $L$ ($R$) modes for individual flavors as well as for the 
  medium.\par
 
 The present manuscript has been organized as follows:
 In section~\ref{two}, we have discussed the quasi-particle model 
 and thermal mass of the quarks in the thermal and magnetic
 field background obtained
  using the perturbative thermal QCD. 
  In subsection~\ref{three1} and~\ref{three2}, we have calculated the Seebeck coefficient 
 without and with magnetic field background, respectively. 
  We disccussed the results in section~\ref{four} 
  and finally we conclude in section~\ref{five}.   

%%%%%%%%%%%%%%%%%%%%%%%%%%%%%%%%%%%%%%%%%%%%%%%%%%%%%%%
\section{Quasi-particle model}\label{two}
%%%%%%%%%%%%%%%%%%%%%%%%%%%%%%%%%%%%%%%%%%%%%%%%%%%%
The central 
feature of quasiparticle models is that a strongly
interacting system of massless quarks and gluons 
 can be described in terms of 
the massive, weakly interacting quasiparticles 
originated due to the collective excitations in the medium. 
There are many 
 quasi-particle models 
such as NJL 
and PNJL models 
\cite{Fukushima:PLB591'2004,
Ghosh:PRD73'2006,Abuki:PLB676'2009,Tsai:JPG36'2009},
 which are based on the 
respective effective QCD models, effective fugacity model
\cite{Chandra:PRC76'2007} and  model based on the 
Gribov–Zwanziger approach
~\cite{Su:PRL114'2015,Florkowski:PRC94'2016,
Jaiswal:PLB11'2020}. 
  Such a  kind of  model
 was first proposed by
Goloviznin and Satz~\cite{Goloviznin:ZPC671'1993}
to study the gluonic plasma and then by Peshier et al.~\cite{Peshier:PRD541996,
Peshier:PRD66'2002} 
to study the equation of state of QGP obtained from
lattice QCD at finite temperature.
At the same time, authors in Refs.~\cite{PlumariPRD84'2011,Bluhm:PLB620'2005,
BluhmPRC:76'2007,Bluhm:PRD77'2008} used quasi-particle 
picture to explain the  lattice data by using a suitable
quasiparticle description for QGP with temperature 
and chemical potential dependent masses.
These results suggest that the high-temperature QGP phase
is suitably described by a thermodynamically consistent
quasiparticle model. In the present study, we have used quasiparticle 
model by Bannur~\cite{Bannur:JHEP09'2007} where 
the total effective mass of the $i^{th}$ quark flavor
with bare quark mass $m_{i,0}$
 has been parametarized as~\cite{Bannur:JHEP09'2007,Bannur:EPJC50'2007,
Lata:PRD95'2017} 
\begin{eqnarray}
m_i^2=m_{i,0}^2+\sqrt{2}m_{i,0}m_{i,T}+m_{i,T}^2,
\label{para_massT}
\end{eqnarray}
to explaine the lattice data with finite bare quark
masses. The thermal mass ($m_{i,T}$) of  
the quark in Eq.~\eqref{para_massT} can be calculated using the HTL perturbation theory as \cite{Braaten:PRD45'1992}
\begin{eqnarray}
m_{iT}^2=\frac{g'^2T^2}{6}\left(1+\frac{\mu^2}{\pi^2 T^2}\right),
\label{quarkmassT}
\end{eqnarray}
where $g'=\sqrt{4\pi\alpha_s}$ refers to the coupling
constant which depends on the 
temperature as
 \begin{eqnarray}
\alpha_s (T)=\frac{g'^{2}}{4 \pi}=
\frac{6\pi}{(33-2N_f) \ln \left(\frac{Q}
{\Lambda_{QCD}}\right)},
\label{coupling_T}
\end{eqnarray}
and $Q$ is set at $2 \pi\sqrt{T^2+\frac{\mu^2}{\pi^2}}$.\par

Now, we will include the strong magnetic field
in the quasiparticle description. 
The  quasi particle mass in the 
presence of  strong $B$ can be generalized as
\begin{eqnarray}
m^2_{i,s}=m_{i0}^2+\sqrt{2}m_{i0}m_{iB,T}+m_{iB,T}^2,
\label{para_massTB}
\end{eqnarray}
where $m_{iB,T}$ is the medium dependent quark mass, which is 
obtained from the pole the resummed propagator. We
know from Dyson-Schwinger
equation   
 \begin{eqnarray}
S^{-1}(p_{\parallel})=\gamma^{\mu}p_{\parallel \mu}
-\Sigma(p_{\parallel}),
\label{dysonB}
\end{eqnarray}
where $\Sigma(p_{\parallel})$  
refer to  the  self-energy of the quark at finite 
$T$ and 
strong $B$, which  has been calculated as~\cite{Rath:Eur80'2020} 
 \begin{eqnarray}
\Sigma(p_\parallel)=\frac{g^2|q_iB|}{3\pi^2}
\left[\frac{\pi T}{2m_{i,0}}-\ln(2)
+\frac{7\mu^2 \zeta(3)}{8\pi^2 T^2}-
\frac{31 \mu^4 \zeta(5)}{32\pi^4 T^4}\right]\left[\frac{\gamma^0p_0}{p_\parallel^2}+\frac{\gamma^3p_z}{p_\parallel^2}+\frac{\gamma^0\gamma^5p_z}{p_\parallel^2}+
\frac{\gamma^3\gamma^5p_0}{p_\parallel^2}\right],
\end{eqnarray}
where $g=\sqrt{4\pi\alpha_s}$ is the running 
coupling which depends on $T$, $B$ and  $\mu$ as
\begin{eqnarray}
\alpha_s(\Lambda^2,eB) =\frac{g^2}{4\pi}
=\frac{\alpha_s(\Lambda^2)}{1+
b_1\alpha_s(\Lambda^2)\ln\left(\frac{\Lambda^2}
{\Lambda^2+eB}\right)},
\label{alpha_B}
\end{eqnarray}
with 
\begin{eqnarray}
\alpha_s(\Lambda^2)=\frac{1}{
b_1\ln\left(\frac{\Lambda^2}
{\Lambda_{\overline{MS}}^2}\right)},
\end{eqnarray}
and $\Lambda$ is set at $2\pi \sqrt{T^2+\frac{\mu^2}
{\pi^2}}$ for quarks,
 $b_1=\frac{11N_c-2N_f}{12\pi}$ and
$\Lambda_{\overline{MS}}=0.176$ GeV.

Due to the heat bath and magnetic field,
the Lorentz (boost) and rotational invariance of the 
system get broken. In such a non-trivial background, 
the covariant form of the 
quark-self energy $\Sigma(p_{\parallel})$ 
can be written as 
\cite{Ayala:PRD91'2015,Karmakar:PRD99'2019}
\begin{eqnarray}
\Sigma(p_{\parallel})=A_1\gamma^{\mu}u_{\mu}+A_2\gamma^{\mu}b_{\mu}
+A_3\gamma^{5}\gamma^{\mu}u_{\mu}+A_4\gamma^{5}\gamma^{\mu}b_{\mu}.
\label{self_quark}
\end{eqnarray}
Here  $u^{\mu}(1,0,0,0)$ 
and $b^{\mu}(0,0,0,-1)$ correspond to 
 the heat bath and the magnetic field, respectively.
 $A_1,A_2,A_3$ and $A_4$ refer to the structure functions,
  which are 
  given  in the LLL approximation as~\cite{Rath:Eur80'2020}
\begin{eqnarray}
A_1&=&\frac{1}{4}{\rm Tr}[\Sigma\gamma^{\mu}u_{\mu}]=
\frac{g^2|q_iB|}{3\pi^2}\left[\frac{\pi T}{2m_{i,0}}-\ln{(2)}
+\frac{7\mu^2 \zeta(3)}{8\pi^2 T^2}-
\frac{31 \mu^4 \zeta(5)}{32\pi^4 T^4}\right]
\frac{p_0}{p_{\parallel}^2},\\
A_2&=&-\frac{1}{4}{\rm Tr}[\Sigma\gamma^{\mu}b_{\mu}]=
\frac{g^2|q_iB|}{3\pi^2}\left[\frac{\pi T}{2m_{i,0}}-\ln{(2)}
+\frac{7\mu^2 \zeta(3)}{8\pi^2 T^2}-
\frac{31 \mu^4 \zeta(5)}{32\pi^4 T^4}\right]
\frac{p_z}{p_{\parallel}^2},\\
A_3&=&\frac{1}{4}{\rm Tr}[\gamma^5 \Sigma\gamma^{\mu}u_{\mu}]=-
\frac{g^2|q_iB|}{3\pi^2}\left[\frac{\pi T}{2m_{i,0}}-\ln{(2)}
+\frac{7\mu^2 \zeta(3)}{8\pi^2 T^2}-
\frac{31 \mu^4 \zeta(5)}{32\pi^4 T^4}\right]
\frac{p_z}{p_{\parallel}^2},\\
A_4&=&-\frac{1}{4}{\rm Tr}[\gamma^5\Sigma\gamma^{\mu}b_{\mu}]=-
\frac{g^2|q_iB|}{3\pi^2}\left[\frac{\pi T}{2m_{i,0}}-\ln{(2)}
+\frac{7\mu^2 \zeta(3)}{8\pi^2 T^2}-
\frac{31 \mu^4 \zeta(5)}{32\pi^4 T^4}\right]
\frac{p_0}{p_{\parallel}^2},
\end{eqnarray}
where $\zeta(3)$ and $\zeta(5)$ correspond to
 the Riemann zeta functions.
  We can further cast  the quark self-energy~\eqref{self_quark} 
 using the chirality 
 projection operators as
\begin{eqnarray}
\Sigma(p_{\parallel})=P_R[(A_1-A_2)\gamma^{\mu}u_{\mu}+(A_2-A_1)\gamma^{\mu}b_{\mu}]P_L
+P_L[(A_1+A_2)\gamma^{\mu}u_{\mu}+(A_2+A_1)\gamma^{\mu}b_{\mu}]P_R,
\end{eqnarray}
where $P_R$ and $P_L$ are the right- and left-handed 
chiral projection operators, respectively,
\begin{eqnarray} 
P_R=\frac{(1+\gamma^{5})}{2}\\
P_L=\frac{(1-\gamma^{5})}{2}.
\end{eqnarray}
%The inverse of the resummed quark propagator can be 
% expressed in terms 
%of  $P_R$ and $P_L$ as
%\begin{eqnarray}
%S^{-1}(p_{\parallel})=P_R\gamma^{\mu}X_{\mu}P_L+P_L\gamma^{\mu}Y_{\mu}P_R,
%\label{inverse_prop}
%\end{eqnarray}
%where 
We obtain the resummed quark propagator in terms
 of $P_R$ and $P_L$ from~\eqref{dysonB} 
\begin{eqnarray}
S(p_{\parallel})=\frac{1}{2}\left[P_L\frac{\gamma^{\mu}X_{\mu}}{X^2/2}P_R+P_R\frac{\gamma^{\mu}Y_{\mu}}{Y^2/2}P_L
\right],
\label{eff_prop}
\end{eqnarray}
where
\begin{eqnarray}
\gamma^{\mu}X_{\mu}&=&\gamma^{\mu}p_{\parallel \mu}-
(A_2-A_1)\gamma^{\mu}b_{\mu}-(A_1-A_2)\gamma^{\mu}u_{\mu},\\
\gamma^{\mu}Y_{\mu}&=&\gamma^{\mu}p_{\parallel \mu}-
(A_2+A_1)\gamma^{\mu}b_{\mu}-(A_1+A_2)\gamma^{\mu}u_{\mu}.
\end{eqnarray}
and
\begin{eqnarray}
&&\frac{X^2}{2}=X_1^2=\frac{1}{2}\left[p_0-(A_1-A_2)\right]^2-\frac{1}{2}\left[p_z+(A_2-A_1)\right]^2 , \\
&&\frac{Y^2}{2}=Y_1^2=\frac{1}{2}\left[p_0-(A_1+A_2)\right]^2-\frac{1}{2}\left[p_z+(A_2+A_1)\right]^2
~.\end{eqnarray}
 The static limit~
($p_0 = 0,~p_z \rightarrow 0$) of the 
poles of the  propagator~\eqref{eff_prop} 
(of either $X_1^2$ or $Y_1^2$ ) gives the
mass of the quark as
\begin{eqnarray}
m_{i,B}^2=\frac{g^2|q_iB|}{3\pi^2}\left[\frac{\pi T}
{2m_{i,0}}-\ln{(2)}+\frac{7\mu^2 \zeta(3)}
{8\pi^2 T^2}-\frac{31 \mu^4 \zeta(5)}{32\pi^4 T^4}\right],
\label{massTB}
\end{eqnarray}
which depends on the magnetic field, temperature
and quark chemical potential.\par

{ The effective  quark mass for $i^{th}$ 
flavor in the case of a weak magnetic field can 
be parameterized like
the earlier cases as
\begin{eqnarray}
m^2_{i,w} = m_{i0}^2 + \sqrt{2}m_{i0}m_{i,L/R}+m_{i,L/R}^2,
\end{eqnarray}
where  $m_{i,L/R}$ refers to the 
 thermal mass for the left- or right-handed 
 chiral mode of  $i^{th}$ flavor  
which can be 
evaluated from the  Dyson-Schwinger equation 
 \begin{eqnarray}\label{inverse_prop1}
S^{*-1}(P) &=& 
\slashed{P} - \Sigma(P).
\label{dyson} 	 
\end{eqnarray} 
 Here $\Sigma(P)$ represents the self-energy of the quark in 
  the weakly magnetized thermal medium which can be
  written in the covariant form 
 at finite $T$ and $B$   as \cite{Das:PRD97'2018}
\begin{eqnarray}\label{52}
	\Sigma(P) = -a_1\slashed{P}-
	a_2\slashed{u}-a_3\gamma_5\slashed{u}-
	a_4\gamma_5\slashed{b},
	\label{self_weak}
\end{eqnarray} 
where $a_1, a_2, a_3, a_4$ 
are the structure functions, which can be 
evaluated by taking the appropriate contractions of 
Eq.~\eqref{52} as~\cite{Das:PRD97'2018} 
\begin{eqnarray}\label{A_net}
	 a_1(p_0,|{\bf{p}}|) &=& 
	\frac{m_{th}^2}{|{\bf{p}}|^2}Q_1\Bigg(\frac{p_0}{|{\bf{p}}|}\Bigg),\\
	 a_2(p_0,|{\bf{p}}|) &=&
	- \frac{m_{th}^2}{|{\bf{p}}|}\Bigg[\frac{p_0}
	{|{\bf{p}}|}Q_1\Big(\frac{p_0}{|{\bf{p}}|}\Big)
	-Q_0\Bigg(\frac{p_0}{|{\bf{p}}|}\Bigg)\Bigg],\\
	 a_3(p_0,|{\bf{p}}|) &=& -4g^2C_FM^2\frac{p_z}
	{|{\bf{p}}|^2}Q_1\Bigg(\frac{p_0}{|{\bf{p}}|}\Bigg),\\
	 a_4(p_0,|{\bf{p}}|) &=& -4g^2C_FM^2\frac{1}
	{|{\bf{p}}|}Q_0\Bigg(\frac{p_0}{|{\bf{p}}|}\Bigg),\label{D_net}
\end{eqnarray}
where
\begin{eqnarray}
	 M^2(T,\mu , B) = \frac{|q_iB|}{16\pi^2}
	\left(\frac{\pi T}{2m_{i0}} -\ln{2}  + 
	\frac{7\mu^2\zeta(3)}{8\pi^2T^2}\right),
\end{eqnarray}
and
 $Q_0$ and $Q_1$ are given by
\begin{eqnarray}
	Q_0(t) &=& \frac{1}{2}\ln\left(\frac{t+1}{t-1}\right),\\
	Q_1(t) &=& \frac{t}{2}\ln\left(\frac{t+1}{t-1}\right)-1 = tQ_0(t)-1.
\end{eqnarray}
 Self-energy~\eqref{self_weak} 
 can be written in the basis of right and left-hand chiral
 projection operators as 
 \begin{eqnarray}\label{63}
	\Sigma(P) = -P_R(a_1\slashed{P} + 
	(a_2+a_3)\slashed{u} + 
	a_4\slashed{b})P_L - P_L(a_1\slashed{P} + 
	(a_2-a_3)\slashed{u} -
	 a_4\slashed{b})P_R.
\end{eqnarray}
%We can re-write the inverse fermion propagator~\eqref{inverse_prop}
%using~\eqref{63} as
%\begin{eqnarray}
%	S^{*-1}(P) = \slashed{P} + P_R\left[A'\slashed{P}
%	 + \left(B'+C'\right)\slashed{u} + 
%	 D'\slashed{b}\right]P_L + P_L
%	 \left[A'\slashed{P} + 
%	 \left(B'-C'\right)\slashed{u} - 
%	 D'\slashed{b}\right]P_R,
%\end{eqnarray}
%which can further be simplified as
%\begin{eqnarray}
%S^{*-1}(P) = P_R \slashed{L} P_L + P_L \slashed{R} P_R.
%\label{inve_PR}		
%\end{eqnarray}
%Since $P_{L,R}\gamma^{\mu} = \gamma^{\mu} P_{R,L}$ and 
%$P_L \slashed{P} P_L = P_R \slashed{P} P_R = P_L P_R \slashed{P} = 0$, 
% $\slashed{L}$ and $\slashed{R}$ are given by
%\begin{eqnarray}
%	\slashed{L} &=& (1+A')\slashed{P} +
%	 (B'+C')\slashed{u}
%	 + D'\slashed{b},\\
%	 \slashed{R} &=& (1+A')\slashed{P} +
%	 (B'-C')\slashed{u} - D'\slashed{b}.
%\end{eqnarray} 
We calculate  the effective quark propagator from~\eqref{dyson} 
\begin{eqnarray}
	S^*(P) = \frac{1}{2}\left[ P_L\frac{\slashed{L}}
	{L^2/2}P_R+P_R\frac{\slashed{R}}{R^2/2}P_L\right],
\end{eqnarray}
where 
\begin{eqnarray}
	& L^2 = (1+a_1)^2P^2 + 2(1+a_1)
	(a_2+a_3)p_0-2a_4(1
	+a_1)p_z+ (a_2+a_3)^2-a_4^2,\\
	& R^2 = (1+a_1)^2P^2 + 2(1+a_1)
	(a_2-a_3)p_0+2a_4
	(1+a_1)p_z+ (a_2-a_3)^2-a_4^2.
\end{eqnarray}
Now in order to get the quark thermal 
mass in weakly magnetized thermal 
QCD medium, we take the static limit
($p_0 =0,|{\bf{p}}|\rightarrow 0$) of
 $L^2/2$ and $R^2/2$ modes,\footnote{We have expanded 
 the Legendre functions 
appearing  in the structure functions in power series of 
 $\frac{|\bf{p}|}{p_0}$ and have  
 considered only upto $\mathcal{O}(g^2)$} 
 we get
\begin{eqnarray}
	&\frac{L^2}{2}\arrowvert_{p_0= 0, {|\bf{p}|}\rightarrow 0}
	 = m_{th}^2 + 4g^2 C_F M^2,\\
	&\frac{R^2}{2}|_{p_0= 0, {|\bf{p}|}\rightarrow 0}
	 = m_{th}^2 - 4g^2 C_F M^2.
\end{eqnarray}
The masses of the left- and right-handed modes 
are given by
\begin{align}\label{qaurk_mass}
	&m_{L}^2 = m_{th}^2 + 4g^2 C_F M^2,\\
	&m_{R}^2 = m_{th}^2 - 4g^2 C_F M^2,
\end{align}
 respectively. We will use these medium generated masses in the 
 dispersion relation of the quarks to 
 calculate the Seebeck and Nernst coefficients in the 
 forthcoming sections.

%%%%%%%%%%%%%%%%%%%%%%%%%%%%%%%%%%%%%%%%%%%%%%%%%%%%%%%%%%%%%%%%%%%%
\section{Theromoelectric response of a thermal QCD medium}\label{three}
%%%%%%%%%%%%%%%%%%%%%%%%%%%%%%%%%%%%%%%%%%%%%%%%%%%%%%%%%%%%%%%%%%%%%
 In the kinetic theory approach, the evolution 
of the phase space 
distribution function is
given  by RBTE, which reads as
\begin{eqnarray}
p^{\mu}\frac{\partial f}{\partial x^{\mu}}+q~F^{\rho \sigma}
p_{\sigma}\frac{\partial f}{\partial p^{\rho}}= 
C[f],
\label{rbte}
\end{eqnarray}
where $f=f_{eq}+\delta f$; $\delta f$ is  
small deviation from the equilibrium
 and $F^{\rho \sigma}$
 corresponds to the electromagnetic field 
strength tensor. $C[f]$ corresponds to the collision
integral, which provides 
microscopic input to the RBTE. 
 In general $C[f]$ is non-linear in $f$, but
  Anderson and Witting proposed a simple collision 
  integral, which
is known as RTA 
\ba
 C[f]=-\frac{p^\mu u_\mu}{\tau} \left(f-f_{\rm eq} 
\right),
\label{RT}
\ea
where $\tau$ is the relaxation time.
 The RTA collision 
term violates 
the particle number and 
current conservation.
This shortcoming is the artifact of the 
linearization of the 
collision term otherwise in principle 
the full collision term
respects all the conservation laws. 
Later this shortcoming 
was circumvented by  Bhatnagar, Gross and Krook (BGK) by modifying 
the RTA  as~\cite{Bhatnagar:PRD94'1954,Schenke:PRD73'2006}
\begin{eqnarray}
C[f] = -\frac{p^\mu u_\mu}{\tau} 
\left(f- \frac{n}
{n_{\mathrm eq}} f_{\rm eq} \right),
\label{bgk}
\end{eqnarray}
where  ${n}$ and ${n_{\mathrm eq}}$ are the 
out of equilibrium and equilibrium number densities, respectively.
The collision term~\eqref{bgk} respects the 
conservation of the particle number {\rm i.e.}
\begin{eqnarray}
\int \frac{d^3p}{(2\pi)^3} C[f]=0.
\end{eqnarray}
In what follows, we will  apply the framework discussed herewith   
to examine the thermoelectric response of the thermal
  medium of quarks and gluons with and without external 
 magnetic field background.
 
%%%%%%%%%%%%%%%%%%%%%%%%%%%%%%%%%%%%%%%%%%%%%%%%%%%%%%%
\subsection{Seebeck coefficient in the 
absence of the magnetic field}\label{three1}
%%%%%%%%%%%%%%%%%%%%%%%%%%%%%%%%%%%%%%%%%%%%%%%%%%%%%%%
In this subsection, we will evaluate the Seebeck coefficient of the 
thermal QCD medium composed of $u$, $d$ and $s$ quarks
(and their anti-particles). 
In the presence of 
the thermal gradient, the charge carriers  
 will move from the hotter regions to 
the colder ones. As a result,
a current is induced in the medium  
 which can be written as
\ba
J_{\mu}=\sum_{i} g_i \int 
\frac{d^3p}{(2\pi)^3}~\frac{p_{\mu}}{\omega_i} 
~(q_i\delta f_i(x,p)+\bar{q_i}\delta \bar{f}_i(x,p)),
\label{current_temp}
\ea
where $\delta f_i(x,p)$ ($\delta \bar{f}_i(x,p)$) refers 
to the infinitesimal deviation in the phase space density 
of quarks (anti-quarks) of $i^{th}$ flavor and $g_i$ corresponds to 
the degeneracy factor.\par
The Boltzmann transport equation~\eqref{rbte} 
in the presence of the 
temperature gradient with BGK collision integral 
can be written as 
\ba
\vec{p}.\frac{\partial f_i}{\partial {\vec{r}}}+q_i ~
{\vec {E}.\vec{p}}
 \frac{\partial f_i}{\partial p^0}+
q_i p_0 ~{\vec{E}.}\frac{\partial f_i}{\partial {\vec{ p}}}=
-p^{\mu}u_{\mu}\nu_i \left(f_i-n_in_{\mathrm eq,i}^{-1} f_{\mathrm eq,i}
\right),
\label{rbte_temp}
\ea
where $f_i=f_{\mathrm eq,i}
+\delta f_i$ and
\ba \label{num_den1}
n_i &=& g_i \int \frac{d^3p}{(2\pi)^3} (f_{\mathrm eq,i}
+\delta f_i),\\
n_{\mathrm eq,i}&=& g_i \int \frac{d^3p}{(2\pi)^3}
f_{\mathrm eq,i}.
\label{num_den2}
\ea
$f_{\mathrm eq,i}$ is the Fermi-Dirac distribution
function and $\nu_i$ is the collision frequency, 
which is estimated by
inverse of the relaxation time~\cite{Hosoya:NPB250'1985}
\ba
\tau_i(T) =\frac{1}{5.1T \alpha_s^2 \log \left(\frac{1}{\alpha_s}\right)
\label{tau_B0} 
[1+0.12(2N_f+1)]},  
\ea 
where $\alpha_s$ is the running coupling 
constant~\eqref{coupling_T}.

The RBTE~\eqref{rbte_temp}
can be recast after some simplification as (see Appendix A)
\ba \label{appendix_A}
\delta f_i- g_in_{\mathrm eq,i}^{-1} 
f_{\mathrm eq,i} \int_{p}\delta f_i &=&
\frac{\vec{p}}{\omega_i}.(\omega_i-\mu)\tau_i \left(-\frac{1}{T^2}\right)
f_{\mathrm eq,i}\left( 
1-f_{\mathrm eq,i} \right)\nabla_{\vec{r}}
 T(\vec{r})\nonumber\\
&&+2q_i\beta \tau_i~ \frac{{\vec{E} \cdot \vec{p}}}
{\omega_i} f_{\mathrm eq,i}\left( 
1-f_{\mathrm eq,i} \right),  
\ea
which can be further solved for $\delta f_i$ as
\begin{eqnarray}
\delta f_i=\delta f_i^{(0)}+g_i 
n_{\mathrm eq,i}^{-1} f_{\rm eq,i} 
\int_{p'}\delta f_i^{(0)},
\label{eqoneseven}
\end{eqnarray}
where
\begin{eqnarray}
\delta f_i^{(0)}=\frac{\vec{p}}{\omega_i}.
(\omega_i-\mu)\tau_i 
\left(-\frac{1}{T^2}\right)
f_{\mathrm eq,i}\left( 
1-f_{\mathrm eq,i} \right)\nabla_{\vec{r}} T(\vec{r})
+\frac{2\beta q_i \tau_i}
{\omega_i} {\vec{E}.\vec{p}} ~f_{\mathrm eq,i}(1-f_{\mathrm eq,i}).
\end{eqnarray}
Following the similar steps, $\delta \bar{f}_i$ can be calculated as 
\begin{eqnarray}
\delta \bar{f}_i=\delta \bar{f}_i^{(0)}+g_i 
n_{\mathrm eq,i}^{-1} \bar{f}_{\rm eq,i} 
\int_{p'}\delta \bar{f}_i^{(0)},
\label{eqoneseven}
\end{eqnarray}
where
\begin{eqnarray}
\delta \bar{f}_i^{(0)}=\frac{\vec{p}}{\omega_i}.(\omega_i +\mu)
\tau_i \left(-\frac{1}{T^2}\right)
\bar{f}_{\mathrm eq,i}\left( 
1-\bar{f}_{\mathrm eq,i} \right)\nabla_{\vec{r}} T(\vec{r})
+2\beta \bar{q}_i \tau_i
\frac{\vec{E}.\vec{p}}{\omega_i}  ~\bar{f}_{\mathrm eq,i}(1-\bar{f}_{\mathrm eq,i}).
\end{eqnarray}
Now substituting  $\delta f_i$ and 
$\delta \bar{f}_i$ in the Eq.~\eqref{current_temp} to obtain
the space part of  induced current
due to a single quark flavor, which reads 
\ba
J_{k,i} &=&   q_i g_i \tau_i \int 
\frac{d^3p}{(2\pi)^3}~ 
\bigg [\bigg \{\frac{p_k^2}{\omega_i^2}(\omega_i-\mu) \left(\frac{-1}{T^2}\right)
f_{\mathrm eq,i}\left( 
1-f_{\mathrm eq,i} \right)\nabla_{\vec{r}} T(\vec{r})
 +2\frac{p_k^2}{\omega_i^2}
 q_iE_k\beta f_{\mathrm eq,i}
(1-f_{\mathrm eq,i})\bigg \} \nonumber\\
&& +\frac{g_i}{ n_{\mathrm eq,i}}~
\frac{p_k}{\omega_i} 
f_{\mathrm eq,i}\int_{p'}
\bigg \{\frac{p_k}{\omega_i}(\omega_i-\mu) 
\left(\frac{-1}{T^2}\right)
f_{\mathrm eq,i}\left( 1-f_{\mathrm eq,i} \right)
\nabla_{\vec{r}} T(\vec{r}) 
+2\frac{p_k}{\omega_i} q_iE_k\beta f_{\mathrm eq,i}(1-f_{\mathrm eq,i})
\bigg \}\bigg]\nonumber\\
&& + \bar{q}_i g_i \tau_i \int 
\frac{d^3p}{(2\pi)^3}~ 
\bigg[\bigg \{\frac{p_k^2}{\omega_i^2}(\omega_i +\mu) \left(\frac{-1}{T^2}\right)
\bar{f}_{\mathrm eq,i}\left( 
1-\bar{f}_{\mathrm eq,i} \right)\nabla_{\vec{r}} T(\vec{r})
+2\frac{p_k^2}{\omega_i^2}
 \bar{q}_iE_k\beta \bar{f}_{\mathrm eq,i}
(1-\bar{f}_{\mathrm eq,i})\bigg \} \nonumber\\
&& +\frac{g_i}{ n_{\mathrm eq,i}} ~\frac{p_k}{\omega_i} 
\bar{f}_{\mathrm eq,i}\int_{p'}
\bigg \{\frac{p_k}{\omega_i}(\omega_i +\mu) \left(-\frac{1}{T^2}\right)
\bar{f}_{\mathrm eq,i}\left( 
1-\bar{f}_{\mathrm eq,i} \right)\nabla_{\vec{r}} T(\vec{r})
+2\frac{p_k}{\omega_i}
 \bar{q}_iE_k\beta \bar{f}_{\mathrm eq,i}
(1-\bar{f}_{\mathrm eq,i})
\bigg \}\bigg].   \nonumber\\ 
\label{induced_c}
\ea
In the state of equilibrium, the resultant current  
due to $i^{th}$ quark flavor beocomes zero {\rm i.e.} $\vec{J}_i=0$. Putting the induced 
current~\eqref{induced_c} to zero,
 we get a relation between the thermal gradient in the 
 medium and  electric field as~\footnote{We
have ommited the flavor label $i$ here for simplicity
as we are interested in the Seebeck coefficint due 
to a single  quark flavor. It has been 
 again  taken when the Seebeck coefficient 
of the medium 
is  considered in Eq.~\eqref{tot_current}.}
\begin{eqnarray}
{\vec{E}} &=&\frac{1}{2 Tq}\left(\frac{L_1+L_2}{L_3+L_4}\right) 
~{ \nabla_{\vec{r}}} T(\vec{r}),\nonumber \\
&\equiv& S~{\nabla_{\vec{r}}} T(\vec{r}).
\label{sstate_b0}
\end{eqnarray}
Here $S$ is the Seebeck coefficient, which
is given by
\ba
S= \frac{1}{2 Tq}\left(\frac{L_1+L_2}{L_3+L_4}\right),
\ea 
where 
\begin{eqnarray}
L_1&=&\int \frac{d^3p}{(2\pi)^3}~ 
\bigg \{\frac{p^2}{3\omega^2}(\omega -\mu) 
f_{\mathrm eq}\left( 
1-f_{\mathrm eq} \right)+\frac{g}{ n_{\mathrm eq}} 
\frac{p}{\omega} 
f_{\mathrm eq}\int_{p'}\frac{p'}{\omega'}(\omega'-\mu)
f_{\mathrm eq}\left( 
1-f_{\mathrm eq} \right)\bigg \},\\
L_2&=&-\int \frac{d^3p}{(2\pi)^3}~ 
\bigg \{\frac{p^2}{3\omega^2}(\omega +\mu) 
\bar{f}_{\mathrm eq}\left( 
1-\bar{f}_{\mathrm eq} \right)+\frac{g}{ \bar{n}_{\mathrm eq}} 
\frac{p}{\omega} 
\bar{f}_{\mathrm eq}\int_{p'}\frac{p'}{\omega'}(\omega' + \mu)
\bar{f}_{\mathrm eq}\left( 
1-\bar{f}_{\mathrm eq} \right)\bigg \},\\
L_3&=&\int \frac{d^3p}{(2\pi)^3}~ 
\bigg \{\frac{p^2}{3\omega^2} 
f_{\mathrm eq}\left( 
1-f_{\mathrm eq} \right)+\frac{g}{ n_{\mathrm eq}} 
\frac{p}{\omega} 
f_{\mathrm eq}\int_{p'}\frac{p'}{\omega'}
f_{\mathrm eq}\left( 
1-f_{\mathrm eq} \right)\bigg \},\\
L_4&=&\int \frac{d^3p}{(2\pi)^3}~ 
\bigg \{\frac{p^2}{3\omega_i^2} 
\bar{f}_{\mathrm eq}\left( 
1-\bar{f}_{\mathrm eq} \right)+\frac{g}{ \bar{n}_{\mathrm eq}}
\frac{p}{\omega} 
\bar{f}_{\mathrm eq}\int_{p'}\frac{p'}{\omega'}
\bar{f}_{\mathrm eq}\left( 
1-\bar{f}_{\mathrm eq} \right)\bigg \}.
\end{eqnarray}
 Upto this point, we have only considered
  a single quark flavor,
we will now focus on 
 the hot QCD medium with multiple quark flavors. In our 
case, we have considered three flavour ($u$,$d$ and $s$ 
quarks and their anti-particles) quark gluon plasma. 
The total induced current  can be written as the 
sum of the currents because of individual flavors as
\ba
\vec{J}&=& \sum_{i}\vec{J}_i \nonumber\\
&=&\left(\frac{q^2_1g_1\tau_1}{T}(L_3+L_4)_1+
\frac{q^2_2g_2\tau_2}{T}(L_3+L_4)_2+....\right)\vec{E}\nonumber\\
&&-\left(\frac{q_1g_1\tau_1}{T^2}(L_1+L_2)_1+
\frac{q_2g_2\tau_2}{T^2}(L_1+L_2)_2+.....\right)
\nabla_{\vec{r}} T(\vec{r}).
\ea
In the steady state condition, the total induced 
current vanishes {\em i.e.} $\vec{J}=0$. As a result, we get
\ba
\vec{E}= \frac{1}{2T}\frac{\sum_i q_ig_i\tau_i(L_1+L_2)_i}
{\sum_i q^2_ig_i\tau_i (L_3+L_4)_i}\nabla_{\vec{r}} 
T(\vec{r}),
\label{tot_current}
\ea
which gives the seebeck coefficient for the 
medium as
\ba
S_{\rm tot}=\frac{1}{2T}\frac{\sum_{i} q_ig_i\tau_i(L_1+L_2)_i}
{\sum_i q^2_ig_i\tau_i (L_3+L_4)_i}.  
\ea
Since all the flavors have same relaxation time and 
degeneracy factor, so the total Seebeck coefficient
for the medium can be expressed in terms of the 
Seebeck coefficient of the individual flavor as
\ba
S_{\rm tot}=\frac{\sum_i S_i q^2_i(L_3+L_4)_i}
{\sum_i q^2_i (L_3+L_4)_i}.  
\ea
In the next subsection, we will explore how the presence
of the background magnetic field modulates the 
thermoelectric response of the hot QCD medium.
%%%%%%%%%%%%%%%%%%%%%%%%%%%%%%%%%%%%%%%%%%%%%%%%%%%%%%%%%%%
%%%%%%%%%%%%%%%%%%%%%%%%%%%%%%%%%%%%%%%%%%%%%%%%%%%%%%%%%%%%
\subsection{Seebeck and Nernst coefficient in the
presence of the magnetic field}\label{three2}
%%%%%%%%%%%%%%%%%%%%%%%%%%%%%%%%%%%%%%%%%%%%%%%%%%%%%%%%%%%%%
%%%%%%%%%%%%%%%%%%%%%%%%%%%%%%%%%%%%%%%%%%%%%%%%%%%%%%%%%%%%
 Now we will calculate the Seebeck coefficient
 in  the magnetic field backgound. 
 Firstly, we will consider the strong field regime,
  where the 
 the energy of the  quark is
 quantized via Landau quantization.
  Then, we will explore the weak field limit, where
 the magnetic field dependence in the transport 
  coefficients enters through the cyclotron 
  frequency, which menifests a classical description of the 
  motion of charged
  particle in the magnetic field.
 %%%%%%%%%%%%%%%%%%%%%%%%%%%%%%%%%%%%%%%%%%%%%%%%%%%%%%%
 %%%%%%%%%%%%%%%%%%%%%%%%%%%%%%%%%%%%%%%%%%%%%%%%%%%%%%%
\subsubsection{The strong magnetic field case}
%%%%%%%%%%%%%%%%%%%%%%%%%%%%%%%%%%%%%%%%%%%%%%%%%%%%
%%%%%%%%%%%%%%%%%%%%%%%%%%%%%%%%%%%%%%%%%%%%%%%%%%%
  In the presence of  
 strong $B$,
 the quark energy gets quantized as~\cite{Gusynin:NPB462'1996}
\ba
\omega_i =\sqrt{p_3^2+m_i^2+2n|q_iB|},  
\ea
where $n=0,1,2....$ correspond to the 
discrete Landau levels and the phase space factor takes 
the form~
\cite{Gusynin:NPB462'1996} 
\ba
 \int \frac{d^3p}{(2\pi)^3} \rightarrow 
 \sum_{n=0}^{\infty}\frac{|q_iB|}{2\pi}
 \int \frac{dp_3}{2\pi} (2-\delta_{n0}).
\ea 
Since we are interested in the strong magnetic field 
limit with scale hierarchy ($|qB|>> T^2>>m_i{^2}$),
 the quarks are confined to 
the lowest Landau levels {\em i.e. $n=0$}.
 A dimensional 
reduction in the quark dynamics takes
place from $3+1$ to $1+1$ dimensions rendering the induced current
along the $z$ direction as
 \begin{eqnarray}
J_3= \sum_{i} g_i \frac{|q_iB|}{4\pi^2}\int {dp_3}~\frac{p_3}{\omega_i} 
~(q_i\delta f_i^{B}+\bar{q}_i\delta \bar{f}_i^{B}),
\label{current_B}
\end{eqnarray}
where $\delta f^{B}_i$ and $\delta \bar{f}^{B}_i$ are the 
 deviations in the quark and anti-quark distribution 
functions, respectively. The RBTE~\eqref{rbte_temp} in the 
strong B becomes
\begin{eqnarray}
p^0\frac{\partial f_i^B}{\partial x^0}+p^3\frac{\partial f_i^B}
{\partial x^3}+q_i F^{03}p_3\frac{\partial f_i^B}{\partial p^0}
+q_iF^{30}p_0\frac{\partial f_i^B}{\partial p^3}=-p^{\mu}u_{\mu}\nu_i^B 
\left(f_i^B-n_i^B{n_{\mathrm eq,i}^{B~-1}} f_{\mathrm eq,i}^B \right),
\label{RBTE_B}
\end{eqnarray}
where $f_i^B=f_{\mathrm eq,i}^B +\delta f_i^B$;
 $f_{\mathrm eq,i}^B$ is given by
\begin{eqnarray}
f_{\mathrm eq,i}^{B}=\frac{1}{e^{\beta \omega_i}+1}.
\end{eqnarray}
Here $\omega_i=\sqrt{p_3^2+m_i^2}$; $m_i$ is the 
quasiparticle mass~\eqref{para_massTB} and $\nu_i^B$ 
 is computed by the inverse of the relaxation time~
\cite{Hattori:PRD95'2017}
\begin{eqnarray}
\tau_i^{B}(p_3;T,|qB|) = \frac{\omega_i (e^{\beta\omega_i}-1)}
{\alpha_s C_Fm_i^2(e^{\beta\omega_i}+1)}
{\left(\int \frac{dp'_3}{\omega_i'
(e^{\beta \omega_i'}+1)}\right)}^{-1}.
\end{eqnarray}
 In Eq.~\eqref{RBTE_B}, 
$n_{\mathrm eq,i}^{B}$ and $n_i^{B}$ are the 
equilibrium and non-equilibrium
number densities of quarks in 
strong $B$, which are given as 
\ba
n_{\mathrm eq,i}^{B}&=&\frac{g_i|q_iB|}{4\pi^2}\int dp_3 ~f_{\rm eq,i}^B,\\
n_i^{B}&=&\frac{g_i|q_iB|}{4\pi^2}\int dp_3 ~(f_{\rm eq,i}^B+\delta f_i^B).
\ea
In order to obtain $\delta f_i^B$, we simplify Eq.~\eqref{RBTE_B} as
\begin{eqnarray}
\delta f_i^B- g_i{n_{\mathrm eq,i}^{B}}^{-1} f_{\rm eq,i}^B \int_{p_3}\delta f_i^B
&=&\tau_i^B \frac{p_3}{\omega_i}(\omega_i-\mu)\left(-\frac{1}{T^2}\right)
f_{\rm eq,i}^{B}\left(1-f_{\rm eq,i}^{B} \right) (\nabla T)_3 \nonumber\\
&&+2q_i\beta \tau_i^B 
\frac{p_3 E_3}{\omega_i} f_{\rm eq,i}^{B}\left( 
1-f_{\rm eq,i}^{B} \right),
\label{deviation}
\end{eqnarray}
which is  further solved for $\delta f_i^B$ upto 
first order in iteration as
\begin{eqnarray}
\delta f_i^B=\delta f_i^{B(0)}+ g_i{n_{\mathrm eq,i}^{B}}^{-1} 
f_{\rm eq,i}^B \int_{p'_{3}}\delta f_i^{B(0)},
\label{eqoneseven}
\end{eqnarray}
where
\begin{eqnarray}
\delta f_i^{B(0)}= \frac{p_3}{\omega_i}
\tau_i^B(\omega_i-\mu)\left(-\frac{1}{T^2}\right)
f_{\rm eq,i}^{B}\left( 
1-f_{\rm eq,i}^{B} \right)(\nabla T)_3+\frac{2q_i\beta \tau_i^B}
{\omega_i} p_3 E_3~f_{\rm eq,i}^{B}(1-f_{\rm eq,i}^{B}).
\end{eqnarray}
Similarly, we can write for the anti-quarks as
\begin{eqnarray}
\delta \bar{f}_i^B=\delta \bar{f}_i^{B(0)}+ g_i{n_{\mathrm eq,i}^{B}}^{-1} 
\bar{f}_{\rm eq,i}^B \int_{p'_{3}}\delta \bar{f}_i^{B(0)},
\label{eqoneseven}
\end{eqnarray}
where
\begin{eqnarray}
\delta \bar{f}_i^{B(0)}=\frac{p_3}{\omega_i}\tau_i^B 
(\omega_i +\mu)\left(-\frac{1}{T^2}\right)
\bar{f}_{\rm eq,i}^{B}\left( 
1-\bar{f}_{\rm eq,i}^{B} \right)(\nabla T)_3+
\frac{2\bar{q}_i\beta \tau_i^B}
{\omega_i} p_3 E_3~\bar{f}_{\rm eq,i}^{B}(1-\bar{f}_{\rm eq,i}^{B}).
\end{eqnarray}
Now substituting $\delta {f}_i^B$ and $\delta \bar{f}_i^B$ in
Eq.~\eqref{current_B} to obtain the induced current 
in the strong magnetic field due to a single $i^{th}$ quark 
flavor 
\ba
J_{3,i}&=& q_i g_i
\frac{ |q_iB|}{4\pi^2} \int {dp_3}\bigg [\bigg \{\frac{p_3^2}
 {\omega_i^2}\tau_i^B
 (\omega_i-\mu)\left(\frac{-1}{T^2}\right)
f_{\rm eq,i}^{B}\left( 
1-f_{\rm eq,i}^{B} \right)(\nabla T)_3
+\frac{p_3^2}{\omega_i^2}~2\beta E_3q_i 
\tau_i^B(p_3) \nonumber\\ 
&&~ \times f_{\rm eq,i}^{B}(1-f_{\rm eq,i}^{B})\bigg \} 
 + \frac{g_i}{{n_{\mathrm eq,i}^{B}}}
 \frac{p_3}{\omega_i}f_{\rm eq,i}^B
\int_{p'_{3}}\bigg \{\frac{p'_3}{\omega_i'}\tau_i^B
 (\omega_i-\mu)\left(\frac{-1}{T^2}\right)
f_{\rm eq,i}^{B}\left( 
1-f_{\rm eq,i}^{B} \right)(\nabla T)_3 \nonumber\\
&&+\frac{p'_3}{\omega_i'}~2\beta E_3q_i\tau_i^B (p_3^\prime)
~f_{\rm eq,i}^{B}~(1-f_{\rm eq,i}^{B})\bigg \}\bigg ] 
 +  \bar{q}_i g_i
\frac{|q_iB|}{4\pi^2} \int {dp_3}\bigg [\bigg \{\frac{p_3^2}
 {\omega_i^2}\tau_i^B
 (\omega_i + \mu)\left(\frac{-1}{T^2}\right)\nonumber \\
&&~ \times \bar{f}_{\rm eq,i}^{B}\left( 
1-\bar{f}_{\rm eq,i}^{B} \right)(\nabla T)_3
+\frac{p_3^2}{\omega_i^2}~2\beta E_3 \bar{q}_i\tau_i^B(p_3)~ 
 \bar{f}_{\rm eq,i}^{B}(1-\bar{f}_{\rm eq,i}^{B})\bigg \}
+\frac{g_i}{{\bar{n}_{\mathrm eq,i}^{B}}} \frac{p_3}
{\omega_i}\bar{f}_{\rm eq,i}^B \nonumber\\
&&~\int_{p'_{3}}\bigg \{\frac{p'_3}{\omega_i'}\tau_i^B
 (\omega_i' +\mu)\left(\frac{-1}{T^2}\right)
\bar{f}_{\rm eq,i}^{B}\left( 
1-\bar{f}_{\rm eq,i}^{B} \right)(\nabla T)_3
 +\frac{p'_3}{\omega_i'}~2\beta E_3\bar{q}_i\tau_i^B (p_3^\prime)
~\bar{f}_{\rm eq,i}^{B}~
(1-\bar{f}_{\rm eq,i}^{B})\bigg \}\bigg ].
\nonumber\\
\label{current_densityB}
\ea
In the state of equilibrium $J_{3,i}$ becomes zero and we get 
the relation (omitting label $i$ for simplicity)
\begin{eqnarray}
E_3&=&\frac{1}{2qT}\left(\frac{H_1+H_2}
{H_3+H_4}\right) (\nabla T)_3,\\
&\equiv&S_B~(\nabla T)_3.
\end{eqnarray}
$S_B$ here corresponds to the Seebeck 
coefficient in the strong $B$ background,
which reads 
\ba
S_B = \frac{1}{2qT}\left(\frac{H_1+H_2}{H_3+H_4}\right),  
\ea
where the integrals $H_1$, $H_2$, $H_3$ and $H_4$  
are given by  
\begin{eqnarray}
H_1&=&\frac{|qB|}{4\pi^2}\int {dp_3} \bigg\{\frac{p_3^2}{\omega^2}\tau^B
 (\omega-\mu)f_{\rm eq}^{B}\left( 1-f_{\rm eq}^{B} \right)
 +g\frac{p_3}{\omega}\frac{f_{\rm eq}^B}{n_{\mathrm eq}^{B}}
\int_{p'_{3}}\frac{p'_3}{\omega'}\tau^B
 (\omega'-\mu)f_{\rm eq}^{B}\left( 
1-f_{\rm eq}^{B} \right)\bigg \},\\
H_2&=& -\frac{|qB|}{4\pi^2}\int {dp_3} 
\bigg\{\frac{p_3^2}{\omega^2}\tau^B
 (\omega+\mu)\bar{f}_{\rm eq}^{B}\left( 1-\bar{f}_{\rm eq}^{B} \right)
 +g\frac{p_3}{\omega}\frac{\bar{f}_{\rm eq}^B}
 {\bar{n}_{\mathrm eq}^{B}}
\int_{p'_{3}}\frac{p'_3}{\omega'}\tau^B
 (\omega'+\mu)\bar{f}_{\rm eq}^{B}\left( 
1-\bar{f}_{\rm eq}^{B} \right)\bigg \},\\
H_3&=& \frac{|qB|}{4\pi^2}\int {dp_3}
 \bigg\{\frac{p_3^2}{\omega^2}\tau^B
f_{\rm eq}^{B}\left( 1-f_{\rm eq}^{B} \right)
 +g\frac{p_3}{\omega}\frac{f_{\rm eq}^B}{n_{\mathrm eq}^{B}}
\int_{p'_{3}}\frac{p'_3}{\omega'}\tau^B
 f_{\rm eq}^{B}\left( 
1-f_{\rm eq}^{B} \right)\bigg \},\\
H_4&=& \frac{|qB|}{4\pi^2}\int {dp_3} 
\bigg\{\frac{p_3^2}{\omega^2}\tau^B
 \bar{f}_{\rm eq}^{B}\left( 1-\bar{f}_{\rm eq}^{B} \right)
 +g\frac{p_3}{\omega}\frac{\bar{f}_{\rm eq}^B}
 {\bar{n}_{\mathrm eq}^{B}}
\int_{p'_{3}}\frac{p'_3}{\omega'}\tau^B
 \bar{f}_{\rm eq}^{B}\left( 
1-\bar{f}_{\rm eq}^{B} \right)\bigg \}.
\end{eqnarray}
For the hot QCD medium consisting of $u$, $d$
and $s$ quarks,
the $3^{\rm rd}$ component of the induced  current 
can be written as the vector sum of the individual
currents as
\ba 
J_3 &=& \sum_{i}J_{3,i}\\
&=&\left(\frac{q^2_1g_1|q_1B|}{T}(H_3+H_4)_1+
\frac{q^2_2g_2 |q_2B|}{T}(H_3+H_4)_2+.....\right)E_3\nonumber\\
&&-\left(\frac{q_1g_1|q_1B|}{T^2}(H_1+H_2)_1+
\frac{q_2g_2|q_2B|}{T^2}(H_1+H_2)_2+....\right)(\nabla T)_3,
\ea
which vanishes in the steady state {\em i.e.} $J_3=0$ and gives
\ba
E_3= \frac{1}{2T}\frac{\sum_i q_ig_i(H_1+H_2)_i}
{\sum_i q^2_ig_i (H_3+H_4)_i}(\nabla T)_3.
\ea
We extract  the Seebeck coefficient for the 
composite medium as
\ba
S^{B}_{\rm tot}=\frac{1}{2T}\frac{\sum_i q_i|q_iB|(H_1+H_2)_i}
{\sum_i q^2_i|q_iB| (H_3+H_4)_i},  
\ea
which  can be  expressed in terms of the
 Seebeck coefficient of the single quark flavor as  
\ba
S^{B}_{\rm tot}=\frac{\sum_i S_i |q_i|^3(H_3+H_4)_i}
{\sum_i |q_i|^3 (H_3+H_4)_i}.   
\ea
In the strong magnetic field, there is no current 
in the transverse direction due
to the one-dimensional (LLL) quark dynamics. 
Hence the Nernst coefficient 
which measures the thermoelectric response in the transverse 
direction, vanishes. In the next subsection, we will explore the 
weak magnetic field regime, where the quark dynamics is not 
affected by the Landau quantization, rather the magnetic 
field dependence enters via the cyclotron frequency,
which manifests the semi-classical description. 
In this scenario, the Nernst coefficient
would also appear.  
\subsubsection{ The weak magnetic field case}
 In the weak magnetic field, the dispersion relation of the 
charged particle is not directly affected by the 
 magnetic field rather $B$ acts as a perturbation.
 So the 1+1 dimensional Landau level kinematics 
 is not applicable.
 The induced four current in the medium is given by
\begin{eqnarray}
J_{\mu}=\sum_{i}g_i \int \frac{d^3p}
{(2\pi)^3}\frac{p_{\mu}}{\epsilon_i}
[q_i\delta f_i+\bar{q}_i\delta \bar{f}_i],
\label{current_weak}
\end{eqnarray}
where $\epsilon_i=\sqrt{p^2+m_i^2}$. 
The RBTE~\eqref{rbte} in the presence of the 
Lorenz force can be written as (see the Appendix)
\begin{eqnarray}
\frac{\partial f_i}{\partial t}+\vec{v}.\frac{\partial f_i}{\partial \vec{r}}+
\vec{F}. \frac{\partial f_i}{\partial \vec{p}}&=&
-\frac{1}{\tau_i}\left(f_i-\frac{n_i}{n_{0,i}}f_{0,i} \right),
\label{rbte_lorenz}
\end{eqnarray}
where $\vec{F}=q_i~(\vec{E}+\vec{v}\times \vec{B})$. Since magnetic 
field is not a dominant energy scale here, we have   
used  the same relaxation time calculated in 
the absence of $B$~\eqref{tau_B0}.  
The magnetic 
field and chemical potential
 dependence in $\tau_i$ will enter through 
 the strong coupling contant~\eqref{alpha_B}.
$n_i$ and $n_{0,i}$ are given by 
Eqs.~\eqref{num_den1} and~\eqref{num_den2},
respectively except the mass in 
the dispersion relation will
be replaced by the thermo-magnetic 
mass calculated in weak magnetic field
in section 3.\par 
Without the loss of generality, 
let us consider the electric field in the $xy$ plane {\em i.e.}
$\vec{E}=E_x \hat{x}+E_y \hat{y}$
and magnetic field in $z$ direction {i.e} $\vec{B}=B\hat{z}$. Then
for QCD medium, which is homogeneous in time, Eq.~\eqref{rbte_lorenz}
takes the form 
\begin{eqnarray}
q_iB\tau_i \left(v_x\frac{\partial f_i}{\partial p_y}-
v_y \frac{\partial f_i}{\partial p_x}\right)-\tau_i
\vec{v} .\frac{\partial f_i}{\partial \vec{r}}-
\tau_i q_i \vec{E} .\frac{\partial f_i}{\partial \vec{p}}=
\left(\delta f_i-g_in_{0,i}^{-1}f_{0,i}\int \delta f_i\right),
\end{eqnarray}
which  can be solved for $\delta f_i$ upto first order as
\begin{eqnarray}
\delta f_i =\delta f_i^{(a)}+g_in_{0,i}^{-1}f_{0,i}\int \delta f_i^{(a)},
\label{rbte_bgk}
\end{eqnarray}
where 
\begin{eqnarray}
\delta f_i^{(a)}=q_iB\tau_i \left(v_x\frac{\partial f_i}
{\partial p_y}-
v_y \frac{\partial f_i}{\partial p_x}\right)-\tau_i
\vec{v} .\frac{\partial f_i}{\partial \vec{r}}-
\tau_i q_i\vec{E} .\frac{\partial f_i}{\partial \vec{p}}.
\end{eqnarray}
In order to solve Eq.~\eqref{rbte_bgk}, 
we take an ansatz~\cite{Feng:PRD96'2017}
\begin{eqnarray}
\delta f_i=\delta f_i^{(b)}+g_in_{0,i}^{-1}
f_{0,i}\int_{p'} \delta f_i^{(b)},
\label{ansatz}
\end{eqnarray}
where 
\ba
\delta f_i^{(b)}=f_i-f_{0,i}= -\tau_i q_i\vec{E} 
.\frac{\partial f_{0,i}}{\partial \vec{p}}
-\vec{\lambda}. \frac{\partial f_{0,i}}{\partial \vec{p}}.
\ea
Equating  Eqs.~\eqref{rbte_bgk} and \eqref{ansatz},
we get
\begin{eqnarray}
\vec{\lambda}. \frac{\partial f_{0,i}}{\partial \vec{p}}-
q_iB\tau_i \left(v_y\frac{\partial f_i}{\partial p_x}-
v_x \frac{\partial f_i}{\partial p_y}\right)-
\tau_i \vec{v} .\frac{\partial f_i}{\partial \vec{r}}=
g_in_{0,i}^{-1}f_{0,i}\left(\int_{p'} \delta f_i^{(a)}+\int_{p'} 
\delta f_i^{(b)}\right).
\label{ansatz_rbte}
\end{eqnarray}
We calculate $\frac{\partial f_i}{\partial p_y}$ and
$\frac{\partial f_i}{\partial p_x}$ using the 
ansatz~\eqref{ansatz} as
\begin{eqnarray}
\left(v_x\frac{\partial f_i}{\partial p_y}-
v_y \frac{\partial f_i}{\partial p_x}\right)=
\bigg(v_y \lambda_x + v_y \tau_i q_i E_x-
v_x \lambda_y -v_x \tau_i q_i E_y\bigg)\frac{\partial f_{0,i}}
{\partial \epsilon}\frac{1}{\epsilon_i},
\label{velocity_term}
\end{eqnarray}
where we have retained terms which are linear in the 
velocity only. 

Now substituting Eq.~\eqref{velocity_term} 
in~\eqref{ansatz_rbte} and after doing some re-arrangement,
we get 
\begin{eqnarray}
v_x\left[\frac{\lambda_x}{\tau_i}-\omega_c \tau_i q_i E_y-
\omega_c \lambda_y +\left(\frac{\epsilon_i -\mu}{T}\right)\frac{\partial T}
{\partial x}\right]+v_y\left[\frac{\lambda_y}{\tau_i}+\omega_c \tau_i q_i E_x+
\omega_c \lambda_x +\left(\frac{\epsilon_i -\mu}{T}\right)\frac{\partial T}
{\partial y}\right]\nonumber\\
+\frac{g_iT}{n_{0,i} \tau_i}\left(\int_{p'} 
\delta f_i^{(a)}+\int_{p'} \delta f_i^{(b)}\right)=0,
\end{eqnarray}
where $\omega_c=\frac{qB}{\epsilon}$ is the cyclotron
frequency. Equating the coefficient of $v_x$ and $v_y$ 
on both side gives  
\begin{eqnarray}\label{lambda1}
\frac{\lambda_x}{\tau_i}-\omega_c \tau_i q_i E_y-
\omega_c \lambda_y +\left(\frac{\epsilon_i -\mu}{T}\right)
\frac{\partial T}
{\partial x}=0,\\
\frac{\lambda_y}{\tau_i}+\omega_c \tau_i q_i E_x+
\omega_c \lambda_x +\left(\frac{\epsilon_i -\mu}{T}\right)
\frac{\partial T}
{\partial y}=0.
\label{lambda2}
\end{eqnarray}
We solve above Eqs.~\eqref{lambda1} and~\eqref{lambda2}
to get $\lambda_x$ and $\lambda_y$ as
\begin{eqnarray}
\lambda_x =- \frac{\omega_c^2 \tau_i^3 }
{1+\omega_c^2 \tau_i^2}q_i E_x -
\frac{\tau_i}{1+\omega_c^2 \tau_i^2}\left(\frac{\epsilon_i -\mu}{T}\right)
\frac{\partial T}{\partial x}+
\frac{\omega_c \tau_i^2 }{1+\omega_c^2 \tau_i^2}q_i E_y-
\frac{\omega_c \tau_i^2 }{1+\omega_c^2 \tau_i^2}\left(\frac{\epsilon_i -\mu}{T}\right)
\frac{\partial T}{\partial y},\\
\lambda_y =- \frac{\omega_c \tau_i^2 }
{1+\omega_c^2 \tau_i^2}q_i E_x+
\frac{\omega_c \tau_i^2}{1+\omega_c^2 \tau_i^2}\left(\frac{\epsilon_i -\mu}{T}\right)
\frac{\partial T}{\partial x}-
\frac{\omega_c^2 \tau_i^3}{1+\omega_c^2 \tau_i^2}q_i E_y-
\frac{ \tau_i}{1+\omega_c^2 \tau_i^2}\left(\frac{\epsilon_i -\mu}{T}\right)
\frac{\partial T}{\partial y}.
\end{eqnarray}
Now we substitute $\lambda_x$ and $\lambda_y$ in 
~\eqref{ansatz} to obtain $\delta f$ which reads
\begin{eqnarray}
\delta f_i = \delta f_i^{(b)}+g_in_{0,i}^{-1}f_{0,i} \int_{p'} \delta f_i^{(b)}, 
\end{eqnarray}
where
\begin{eqnarray}
\delta f_i^{(b)}&=&\frac{\partial f_{0,i}}{\partial \epsilon}\left[-\frac{\tau_i}
{1+\omega_c^2 \tau_i^2}q_i v_x+
\frac{\omega_c \tau_i^2 }{1+\omega_c^2 \tau_i^2}q_i v_y\right]E_x\nonumber\\
&&+\frac{\partial f_0}{\partial \epsilon}\left[-\frac{\tau_i}
{1+\omega_c^2 \tau_i^2}q_i v_y
 -\frac{\omega_c \tau_i^2 }{1+\omega_c^2 \tau_i^2}q_i v_x\right]E_y\nonumber\\
&&+\frac{\partial f_0}{\partial \epsilon}\left[\frac{ \tau_i}
{1+\omega_c^2 \tau_i^2}\left(\frac{\epsilon_i -\mu}{T}\right) v_x -
\frac{\omega_c \tau_i^2 }{1+\omega_c^2 \tau_i^2}
\left(\frac{\epsilon_i -\mu}{T}\right)v_y\right]\frac{\partial T}{\partial x}\nonumber\\
&& +\frac{\partial f_{0,i}}{\partial \epsilon}\left[\frac{ \tau_i}
{1+\omega_c^2 \tau_i^2}\left(\frac{\epsilon_i -\mu}{T}\right) v_y +
\frac{\omega_c \tau_i^2 }{1+\omega_c^2 \tau_i^2}
\left(\frac{\epsilon_i -\mu}{T}\right)v_x\right]\frac{\partial T}{\partial y}.
\label{delta_f1}
\end{eqnarray}
Similarly, deviation in the anti-quark distribution function
can be evaluated as (replacing $q_i$ with -$q_i$ and $\omega_c$ with 
 - $\omega_c$ in Eq.~\eqref{delta_f1})
\begin{eqnarray}
\delta \bar{f}_i = \delta \bar{f}_i^{(b)}
+g_i\bar{n}_{0,i}^{-1}
\bar{f}_{0,i} \int_{p'} \delta \bar{f}_i^{(b)} 
\end{eqnarray}
where
\begin{eqnarray}
\delta \bar{f}_i^{(b)}&=&\frac{\partial 
\bar{f}_{0,i}}{\partial \epsilon}\left[\frac{\tau_i}
{1+\omega_c^2 \tau_i^2}q_i v_x+
\frac{\omega_c \tau_i^2 }{1+\omega_c^2 
\tau_i^2}q_i v_y\right]E_x\nonumber\\
&&+\frac{\partial \bar{f}_{0,i}}{\partial 
\epsilon}\left[\frac{\tau_i}
{1+\omega_c^2 \tau_i^2}q_i v_y
 -\frac{\omega_c \tau_i^2 }{1+\omega_c^2 
 \tau_i^2}q_i v_x\right]E_y\nonumber\\
&&+\frac{\partial \bar{f}_{0,i}}{\partial 
\epsilon}\left[\frac{ \tau_i}
{1+\omega_c^2 \tau_i^2}\left(\frac{\epsilon_i -\mu}{T}\right) v_x +
\frac{\omega_c \tau_i^2 }{1+\omega_c^2 \tau_i^2}
\left(\frac{\epsilon_i -\mu}{T}\right)v_y\right]
\frac{\partial T}{\partial x}\nonumber\\
&& +\frac{\partial \bar{f}_{0,i}}{\partial \epsilon}
\left[\frac{ \tau_i}
{1+\omega_c^2 \tau_i^2}\left(\frac{\epsilon_i -\mu}{T}
\right) v_y -
\frac{\omega_c \tau_i^2 }{1+\omega_c^2 \tau_i^2}
\left(\frac{\epsilon_i -\mu}{T}\right)v_x\right]
\frac{\partial T}{\partial y}.
\end{eqnarray}
 We substitute  $\delta f_i$ and  $\delta \bar{f}_i$
in~\eqref{current_weak} to get the $x$ and $y$
components of the induced current density 
due to $i^{th}$ quark flavor 
\begin{eqnarray}\label{J_x}
J_{x,i}&=&q_ig_i\left[(q_i\beta I_{1,i})E_x+
(q_i\beta I_{2,i})E_y
+(\beta^2 I_{3,i})\frac{\partial T}{\partial x}+
(\beta^2 I_{4,i})\frac{\partial T}{\partial y}\right],\\
J_{y,i}&=&q_ig_i\left[(-q_i\beta I_{2,i})E_x
+(q_i\beta I_{1,i})E_y
+(-\beta^2 I_{4,i})\frac{\partial T}{\partial x}+
(\beta^2 I_{3,i})\frac{\partial T}{\partial y}
\right].
\label{J_y}
\end{eqnarray}
The integrals 
$I_1$, $I_2$, $I_3$ and $I_4$ in the above equations 
 are given by (omitting  
label $i$ for simplicity)
\ba
I_1=I_{1q}+I_{1\bar{q}}, \\
I_2=I_{2q}+I_{2\bar{q}},\\
I_3=I_{3q}+I_{3\bar{q}},\\
I_4=I_{4q}+I_{4\bar{q}},
\ea
where
\begin{eqnarray*}
I_{1q}&=&\int \frac{d^3p}{(2\pi)^3}~\left \{\frac{p^2}{3\epsilon^2}
\frac{ \tau}{(1+\omega_c^2 \tau^2)}
f_0(1-f_0)+\frac{g}{n_0}\frac{p}
{\epsilon}f_0 \int_{p'}\frac{p'}
{\epsilon'}\frac{\tau}{(1+\omega_c^2 \tau^2)}
f_0(1-f_0) \right \},\\
I_{1\bar{q}}&=&\int \frac{d^3p}{(2\pi)^3}~\left \{\frac{p^2}{3\epsilon^2}
\frac{ \tau}{(1+\omega_c^2 \tau^2)}
\bar{f}_0(1-\bar{f}_0)+\frac{g}{\bar{n}_0}\frac{p}
{\epsilon}f_0 \int_{p'}\frac{p'}
{\epsilon'}\frac{\tau}{(1+\omega_c^2 \tau^2)}
\bar{f}_0(1-\bar{f}_0)\right \},\\
I_{2q}&=&\int \frac{d^3p}{(2\pi)^3}~\left \{\frac{p^2}{3\epsilon^2}
\frac{\omega_c \tau^2}{(1+\omega_c^2 \tau^2)}
f_0(1-f_0)+\frac{g}{n_0}\frac{p}
{\epsilon}f_0 \int_{p'}\frac{p'}
{\epsilon'}\frac{\omega_c \tau^2}
{(1+\omega_c^2 \tau^2)}f_0(1-f_0)\right \},\\
I_{2\bar{q}}&=&-\int \frac{d^3p}{(2\pi)^3}~\left \{\frac{p^2}{3\epsilon^2}
\frac{\omega_c \tau^2}{(1+\omega_c^2 \tau^2)}
\bar{f}_0(1-\bar{f}_0)+\frac{g}{\bar{n}_0}\frac{p}
{\epsilon}f_0 \int_{p'}\frac{p'}
{\epsilon'}\frac{\omega_c \tau^2}{(1+\omega_c^2 \tau^2)}
\bar{f}_0(1-\bar{f}_0)\right \},\\
I_{3q}&=&-\int \frac{d^3p}{(2\pi)^3}~\left \{\frac{p^2}{3\epsilon^2}
\frac{\tau}{(1+\omega_c^2 \tau^2)}
(\epsilon -\mu)f_0(1-f_0)+\frac{g}{n_0}\frac{p}
{\epsilon}f_0 \int_{p'}\frac{p'}
{\epsilon'}\frac{\tau}{(1+\omega_c^2 \tau^2)}(\epsilon' -\mu)f_0(1-f_0)\right \},\\
I_{3\bar{q}}&=&\int \frac{d^3p}{(2\pi)^3}~\left \{\frac{p^2}{3\epsilon^2}
\frac{\tau}{(1+\omega_c^2 \tau^2)}
(\epsilon +\mu)\bar{f}_0(1-\bar{f}_0)+\frac{g}{\bar{n}_0}\frac{p}
{\epsilon}f_0 \int_{p'}\frac{p'}
{\epsilon'}\frac{\tau}{(1+\omega_c^2 \tau^2)}(\epsilon' +\mu)
\bar{f}_0(1-\bar{f}_0)\right \},\\
I_{4q}&=&-\int \frac{d^3p}{(2\pi)^3}~\left \{\frac{p^2}{3\epsilon^2}
\frac{\omega_c \tau^2}{(1+\omega_c^2 \tau^2)}(\epsilon -\mu)
f_0(1-f_0)+\frac{g}{n_0}\frac{p}
{\epsilon}f_0 \int_{p'}\frac{p'}
{\epsilon'}\frac{\omega_c \tau^2}{(1+\omega_c^2 \tau^2)}
(\epsilon' -\mu)f_0(1-f_0)\right \},\\
I_{4\bar{q}}&=&-\int \frac{d^3p}{(2\pi)^3}~\left \{\frac{p^2}{3\epsilon^2}
\frac{\omega_c \tau^2}{(1+\omega_c^2 \tau^2)}(\epsilon +\mu)
\bar{f}_0(1-\bar{f}_0)+\frac{g}{\bar{n}_0}\frac{p}
{\epsilon}f_0 \int_{p'}\frac{p'}
{\epsilon'}\frac{\omega_c \tau^2}{(1+\omega_c^2 \tau^2)}
(\epsilon' +\mu)\bar{f}_0(1-\bar{f}_0)\right \}.
\end{eqnarray*}
In the state of equilibrium, the components of the 
induced current density along $x$ and $y$ direction
vanishes {\em i.e.} $J_{x,i}=J_{y,i}=0$. We can write 
from Eqs.~\eqref{J_x} and~\eqref{J_y}
\ba \label{J_x0} 
C_1 E_x+C_2E_y+C_3\frac{\partial T}{\partial x}
+C_4\frac{\partial T}{\partial y}&=&0, \\
-C_2 E_x+C_1E_y-C_4\frac{\partial T}{\partial x}
+C_3\frac{\partial T}{\partial y}&=&0, 
\label{J_y0}
\ea
provided $C_1=qI_1$, $C_2=qI_2$,
 $C_3=\beta I_3$, $C_4=\beta I_4$. Thermoelectric transport coefficients are related to   the electric 
 field components and  temperature
 gradients 
  via a matrix equation 
\ba
  \begin{pmatrix}
  E_x \\
  E_y
  \end{pmatrix}
  =\begin{pmatrix}
  S & N|B|\\
  -N|B| & S
  \end{pmatrix}
  \begin{pmatrix}
  \frac{\partial T}{\partial x}~\\
  \frac{\partial T}{\partial y}
  \end{pmatrix}.
\ea   
We solve Eqs.~\eqref{J_x0}  and \eqref{J_y0} for 
$E_x$ and $E_y$  
 as 
\ba
E_x=\left(-\frac{C_1C_3+C_2C_4}{C_1^2+C_2^2}\right)
\frac{\partial T}{\partial x} 
+\left(-\frac{C_2C_3-C_1C_4}{C_1^2+C_2^2}\right)
\frac{\partial T}{\partial y},\\
E_y=\left(-\frac{C_1C_3+C_2C_4}{C_1^2+C_2^2}\right)
\frac{\partial T}{\partial y} 
-\left(-\frac{C_2C_3-C_1C_4}{C_1^2+C_2^2}\right)
\frac{\partial T}{\partial x},
\ea
which give the Seebeck and Nernst coefficients 
\ba
S &=& -\frac{(C_1C_3+C_2C_4)}{C_1^2+C_2^2},\\
N|B|&=&\frac{(C_2C_3-C_1C_4)}{C_1^2+C_2^2},
\ea
respectively. The integrals $C_2$ and $C_4$ vanishes in the 
absence of the magnetic field, as a result Nernst 
coefficient would also vanish.\par

Now we will compute the Seebeck and 
Nernst coefficients for the medium. In the medium 
composed of $u$ and $d$ light quarks, the 
$x$ and $y$ components of the current can be written as the sum of the individual contributions as
\begin{eqnarray}
J_x&=&\sum_{i=u,d}\left[q_i (I_1)_iE_x+q_i (I_2)_iE_y
+\beta (I_3)_i\frac{\partial T}{\partial x}+
\beta (I_4)_i\frac{\partial T}{\partial y}\right],\\
J_y&=&\sum_{i=u,d}\left[q_i (I_2)_iE_x+q_i (I_1)_iE_y-
\beta (I_4)_i\frac{\partial T}{\partial x}+
\beta (I_3)_i\frac{\partial T}{\partial y}
\right].
\end{eqnarray}
 The 
Seebeck and Nernst coefficients of the medium can be extracted 
by imposing the 
steady state condition ({\em i.e.} putting $J_x=J_y=0$) as
\ba
S^{B'}_{tot} &=& -\frac{(K_1K_3+K_2K_4)}{K_1^2+K_2^2},\\
N|B|&=&\frac{(K_2K_3-K_1K_4)}{K_1^2+K_2^2}.
\ea
where
\ba
K_1&=&\sum_{i=u,d}q_i (I_1)_i, \quad \quad 
 K_2=\sum_{i=u,d}q_i (I_2)_i,\\
K_3&=&\sum_{i=u,d}\beta(I_3)_i, \quad \quad  
K_4=\sum_{i=u,d}\beta (I_4)_i.
\ea
\section{Results and discussion}\label{four}
 In this section, we will discuss the 
results
obtained in the 
previous sections numerically.
 In Fig.~\ref{seebeck_B01} (a), we  display 
 the variation of the Seebeck coefficient
 with $T$ 
 in the BGK and RTA collision terms 
for $u$ quarks at $\mu=60$ MeV. It was found  that the
magnitude of the Seebeck coefficient decreases with $T$
in both the collision terms.
 We have computed the ratio of the Seebeck coefficients 
in BGK collision term to that calculated with the 
RTA (BGK/RTA) to get the numerical estimates of the relative
competition between the two collision integrals. The 
ratio is found to be around $\sim 0.98$ for the 
individual flavors in the temperature domain $160-400$
MeV, which indicates that Seebeck coefficient 
is slightly reduced in the BGK term. The sign of the Seebeck coefficient for $d$ 
and $s$ quarks get reversed
due to their negative charges  
[see Fig.~\ref{seebeck_B01} (b) and Fig.~\ref{seebeck_B02} (a)].
In the case of the composite medium, the Seebeck coefficient
 ($S_{\rm tot}$)
[see Fig.~\ref{seebeck_B02} (b)]  has been 
found to be positive. We notice a considerable enhancement
in the magnitude of $S_{\rm tot}$ in BGK 
collision term, which is around 
$12 \%$ at lower $T$ (160 MeV) and $26 \%$ 
at high $T$ (400 MeV) [see Fig.~\ref{seebeck_B03} (b)].
We further study the effects of the quark chemical 
potential ($\mu$) on the medium Seebeck coefficient  
 in the BGK
collision term in  Fig.~\ref{seebeck_B03} (a) taking
 the strength of   
 $\mu= 40, 60$ and $ 80$ MeV
and have found that  $S_{\rm tot}$
 increases as we raise $\mu$. Similar results
 of $\mu$ dependence in Seebeck coefficient
  have been found 
 in Ref.~\cite{Dey:PRD102'2020} with RTA collision integral. We can conclude that both BGK collision term and baryon 
 asymmetry in the medium enhance its ability 
 to convert the temperature gradient into the current.  \par 

\vspace{8mm}
\begin{center}
\begin{figure}[H]
\begin{tabular}{cc}
\includegraphics[width=6.0cm]{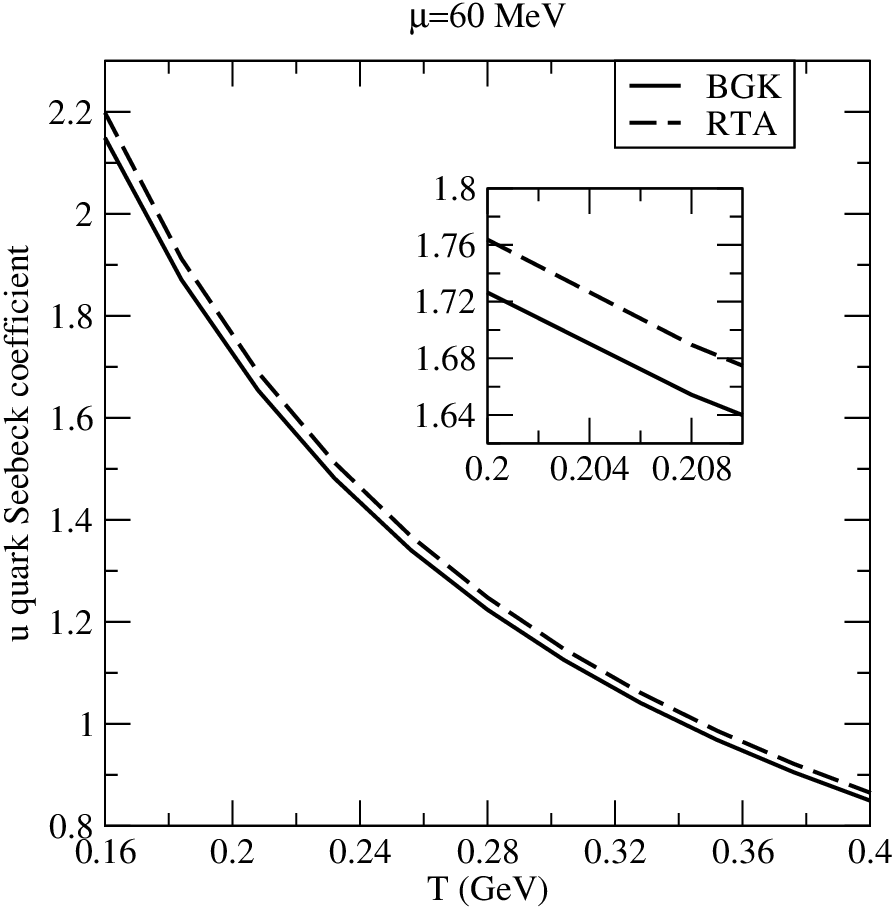}&
\hspace{2.5cm}
\includegraphics[width=6.0cm]{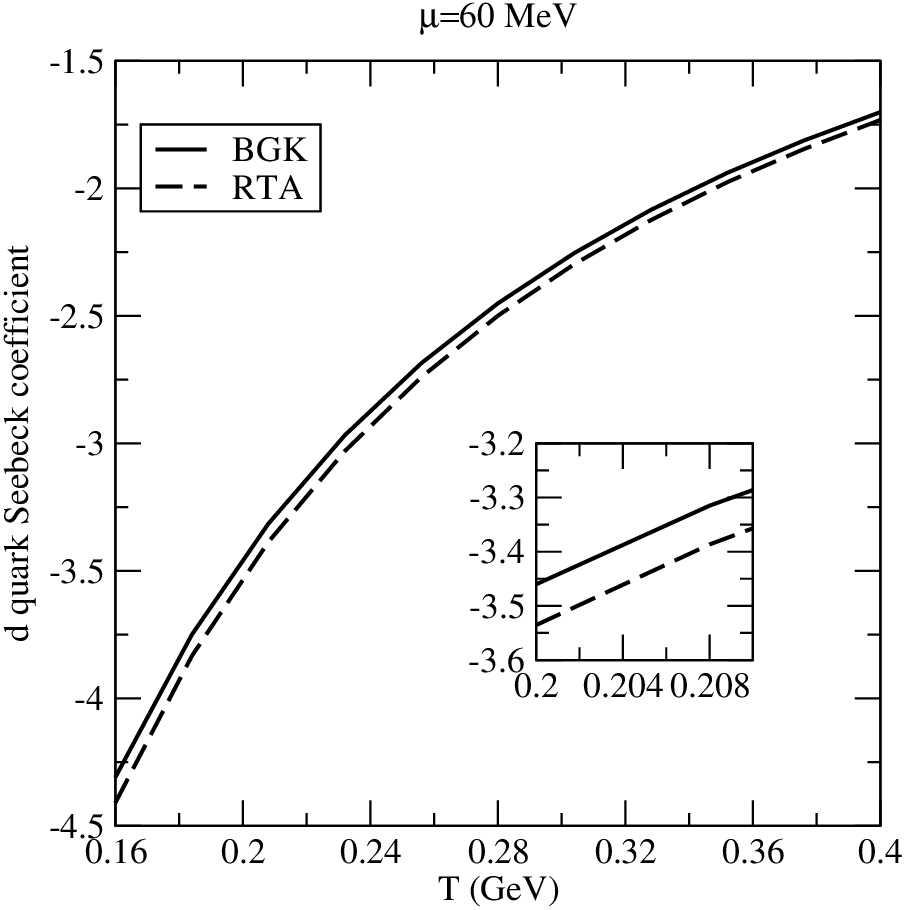}\\
(a)& \hspace{2.5cm} (b) 
\end{tabular}
\caption{Temperature dependence of the Seebeck coefficient in
$B=0$ case: {\bf (a)} for $u$ quarks and
{\bf (b)} for $d$ quarks }
\label{seebeck_B01} 
\end{figure}
\end{center}

\begin{center}
\begin{figure}[H]
\begin{tabular}{cc}
\includegraphics[width=6.0cm]{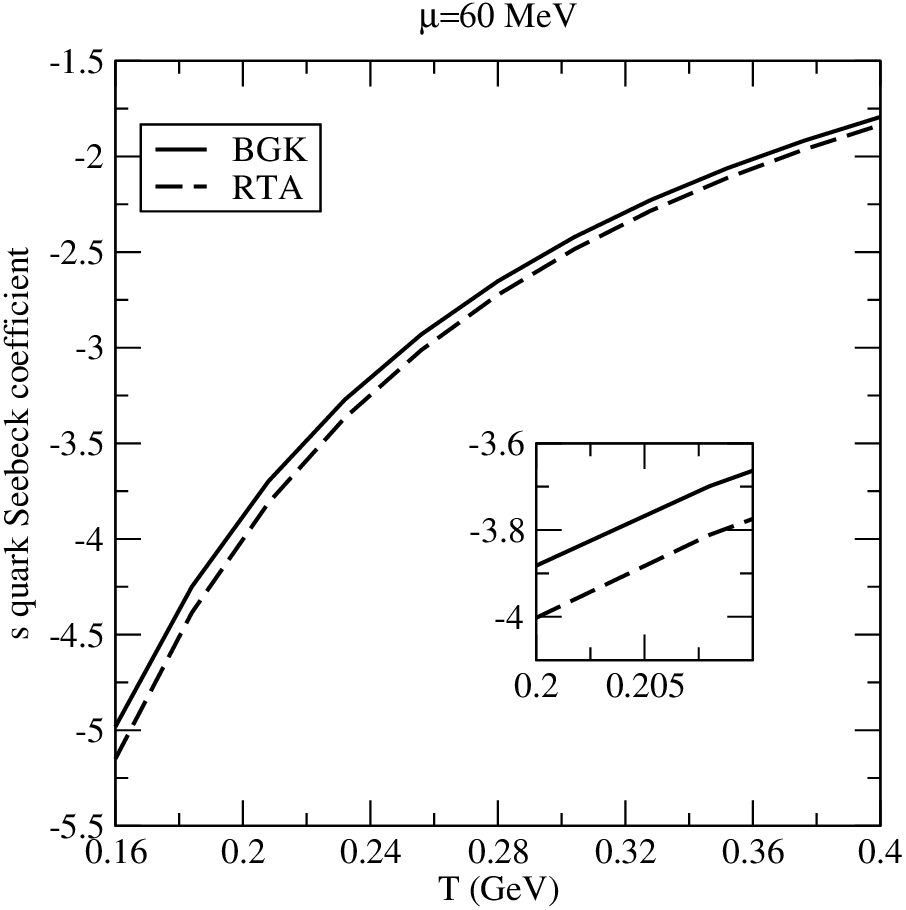}&
\hspace{02.5cm}
\includegraphics[width=6.0cm]{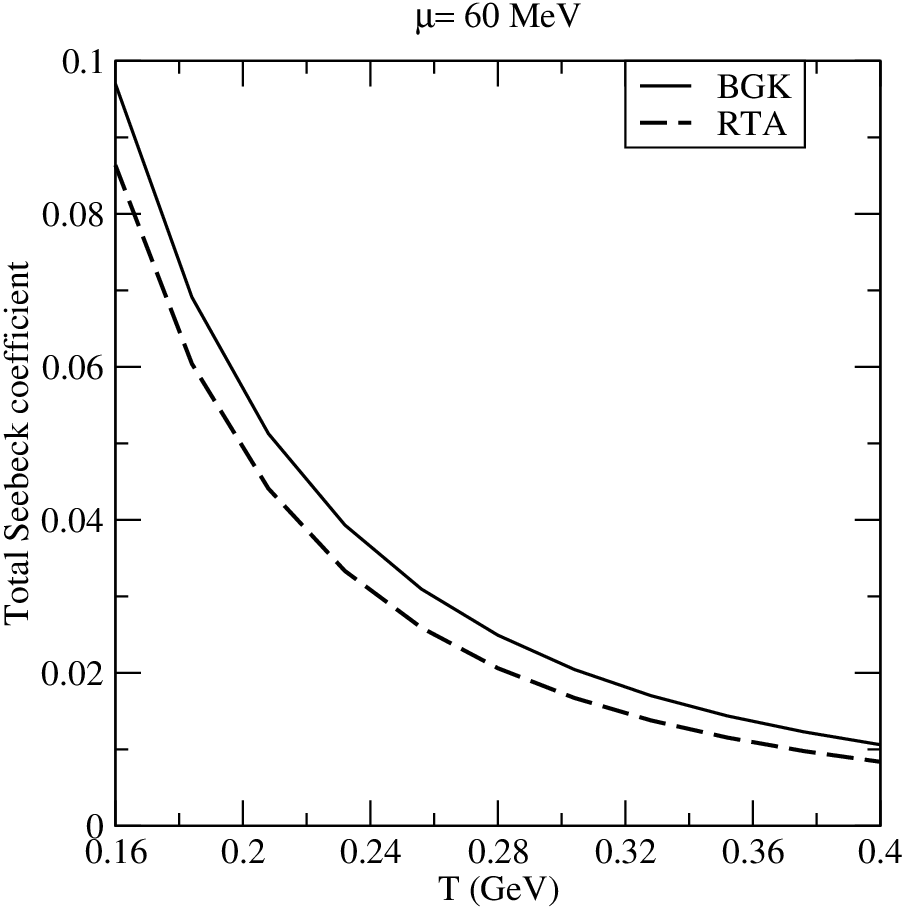}\\
(a)& \hspace{2.5cm} (b) 
\end{tabular}
\caption{Temperature dependence of the Seebeck coefficient 
for $B=0$ case: {\bf (a)} for $s$ quarks and
 {\bf (b)} for composite
medium}
\label{seebeck_B02} 
\end{figure}
\end{center}

\begin{center}
\begin{figure}[H]
\begin{tabular}{cc}
\includegraphics[width=6cm]{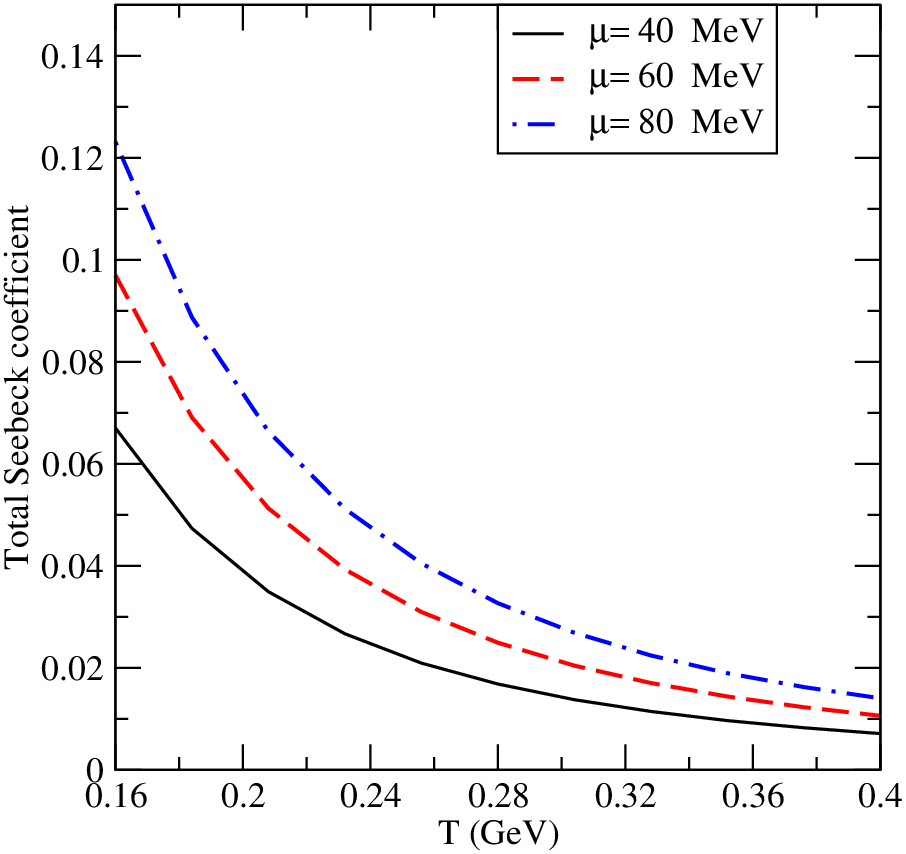}&
\hspace{2.5cm}
\includegraphics[width=6cm]{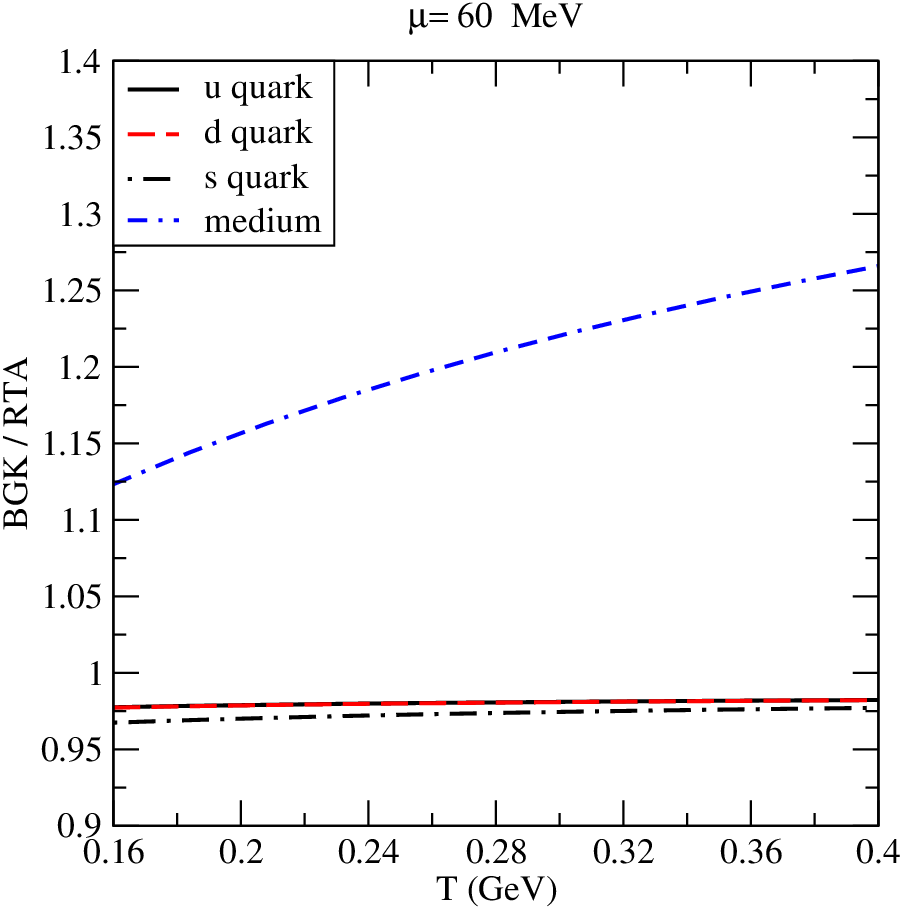}\\
(a) & \hspace{2.5cm} (b) 
\end{tabular}
\caption{{\bf Left panel:} Temperature dependence of
 Seebeck coefficient for the medium with BGK collision term
in absence of  magnetic field for different strengths of
 $\mu$. {\bf Right panel:} Ratio of Seebeck 
 coefficients in BGK to that in RTA collision 
 integral with temperature} 
\label{seebeck_B03} 
\end{figure}
\end{center}

 In Figs.~\ref{seebeck_B1} and \ref{seebeck_B2},
we explore the effects of the BGK 
collision integral on the Seebeck coefficient of a strongly magnetized hot QCD medium. We have chosen the strength
of the magnetic field as $eB=15m_{\pi}^2$ and 
$10m_{\pi}^2$ with $\mu=60$ MeV. 
The Seebeck coefficient for the individual quark
flavors as well 
as for the combined medium gets enhanced in the BGK
collision integral considerably.  
The enhancement 
is around $18 \%- 25 \%$ (for  
$u$ quarks) and $16 \%- 27 \%$ (for $d$ quarks) 
in the temperature range $160 <T< 400$ MeV.
For the $s$ quark, the enhancement is between  $21 \% $ to
$37 \%$ in the same domain of $T$. 
In the case of the medium, it
is around $16 \%$ near $T_c$ but decreases 
as we go towards higher temperatures~
[see Fig.~\ref{seebeck_B3} (b)].
$S^B_{\rm tot}$ increases with the  
strength of $\mu$ like the $B=0$ 
case~[see Fig.~\ref{seebeck_B3} (a)], which 
is in agreement with the study made 
in Ref.~\cite{Dey:PRD104'2021} 
in the RTA framework.  
\par

\vspace{8mm}
\begin{center}
\begin{figure}[H]
\begin{tabular}{cc}
\includegraphics[width=6.0cm]{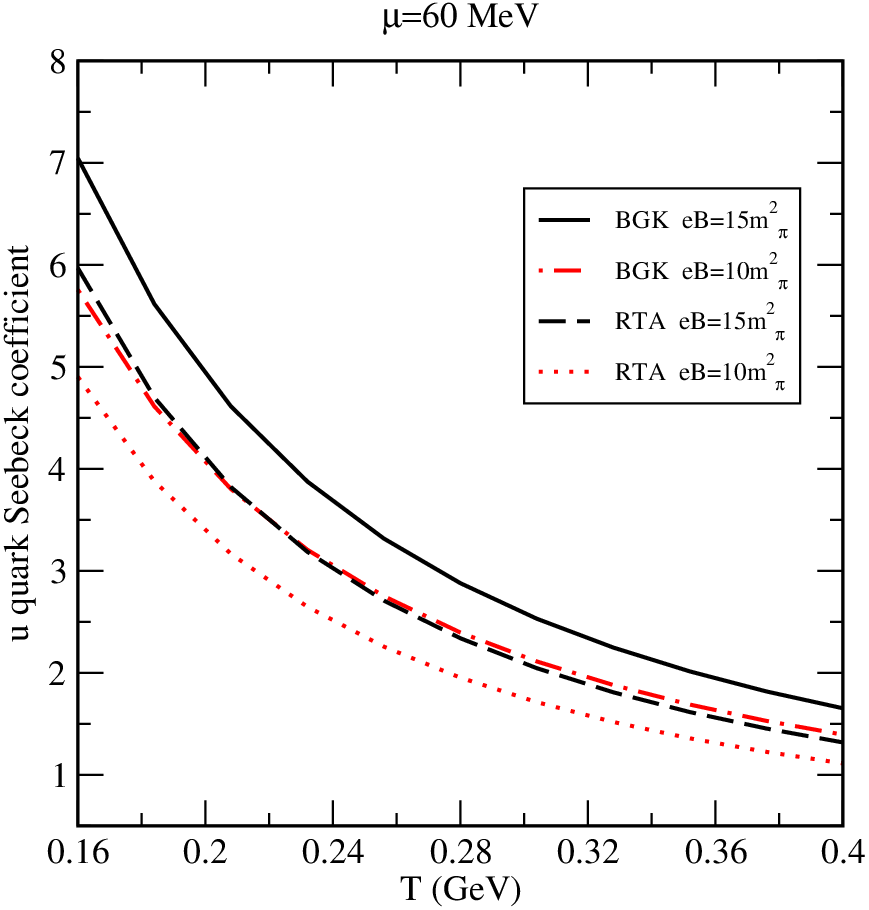}&
\hspace{2.5cm}
\includegraphics[width=6.0cm]{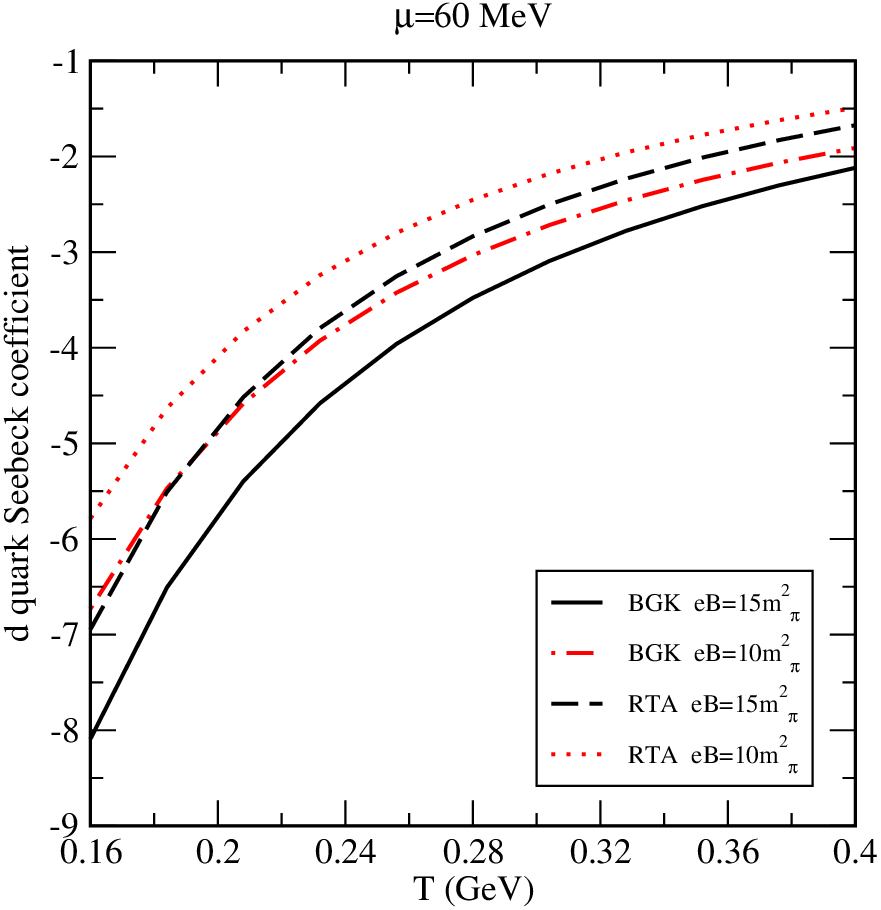}\\
(a) & \hspace{2.5cm} (b)
\end{tabular}
\caption{Temperature dependence of the Seebeck coefficient in
the strong $B$: {\bf (a)} for $u$ quarks
and {\bf (b)} for $d$ quarks }
\label{seebeck_B1} 
\end{figure}
\end{center}

\begin{center}
\begin{figure}[H]
\begin{tabular}{cc}
\includegraphics[width=6.0cm]{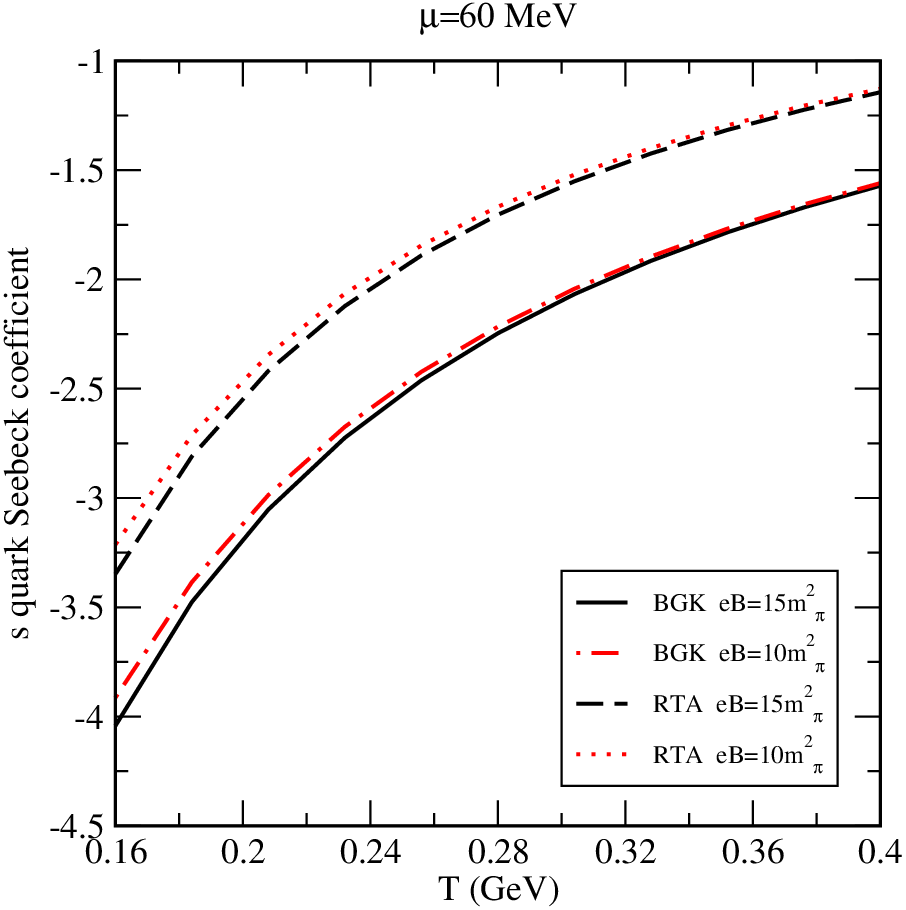}&
\hspace{2.5cm}
\includegraphics[width=6.0cm]{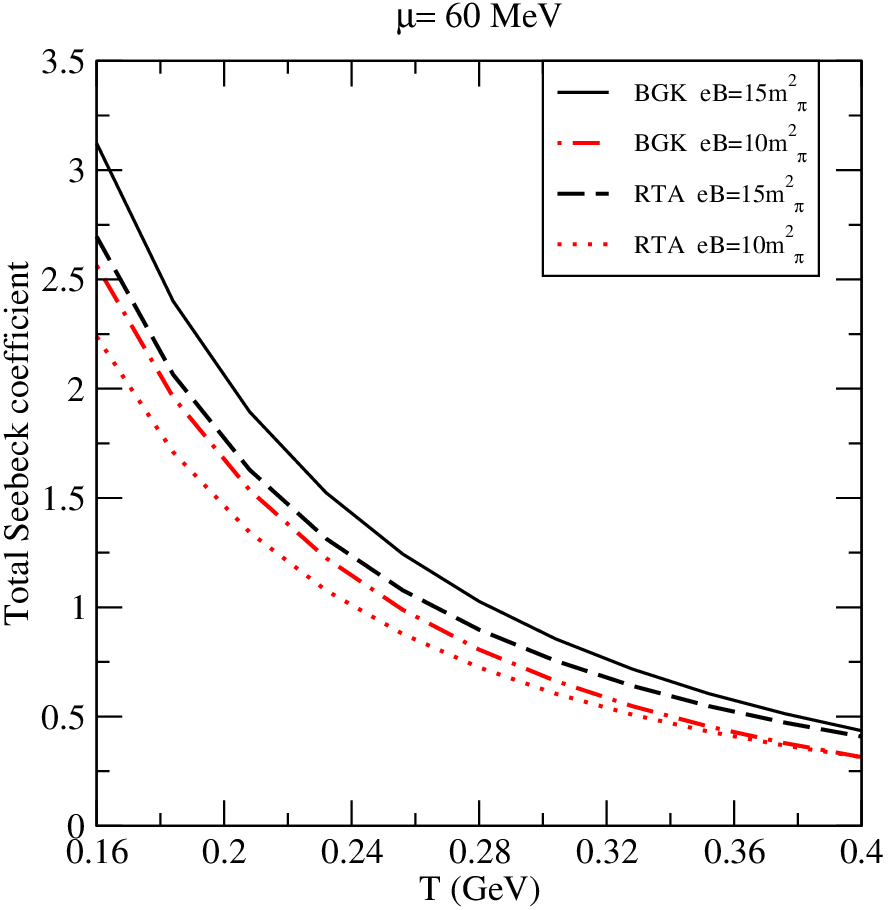}\\
(a) & \hspace{2.5cm} (b)
\end{tabular}
\caption{Temperature dependence of the Seebeck coefficient in
strong $B$: {\bf (a)} for $s$ quarks
 and {\bf (b)} for composite
medium}
\label{seebeck_B2} 
\end{figure}
\end{center}

\begin{center}
\begin{figure}[H]
\begin{tabular}{cc}
\includegraphics[width=6cm]{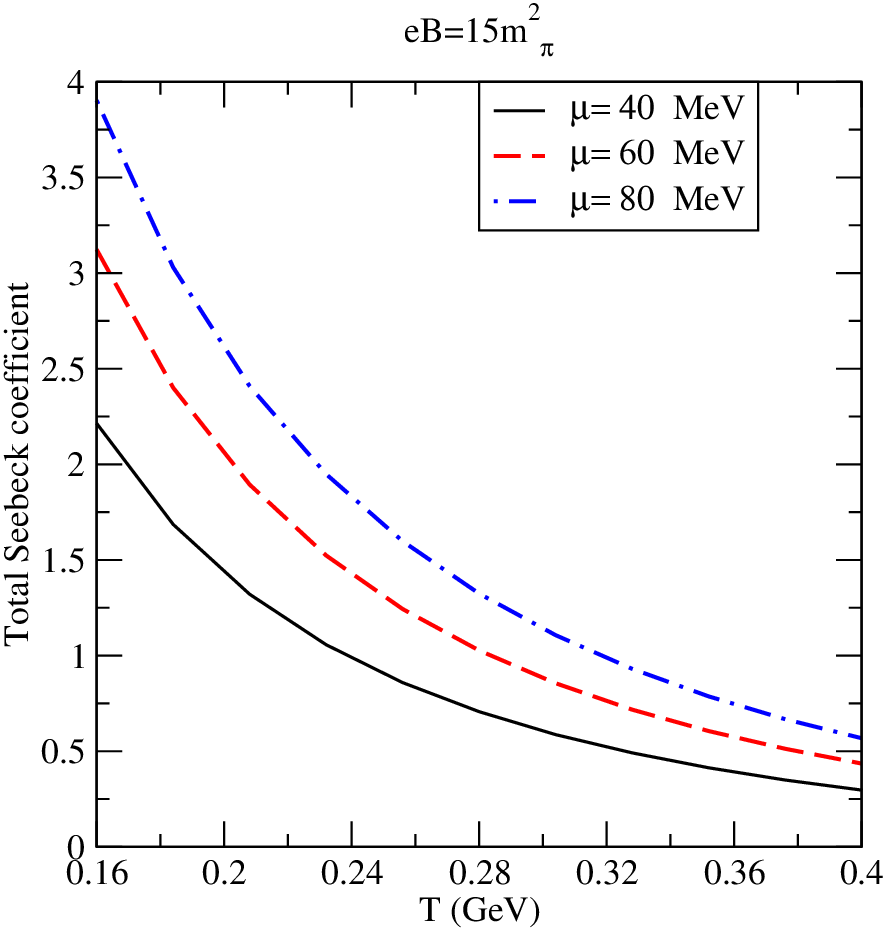}&
\hspace{2.5cm}
\includegraphics[width=6cm]{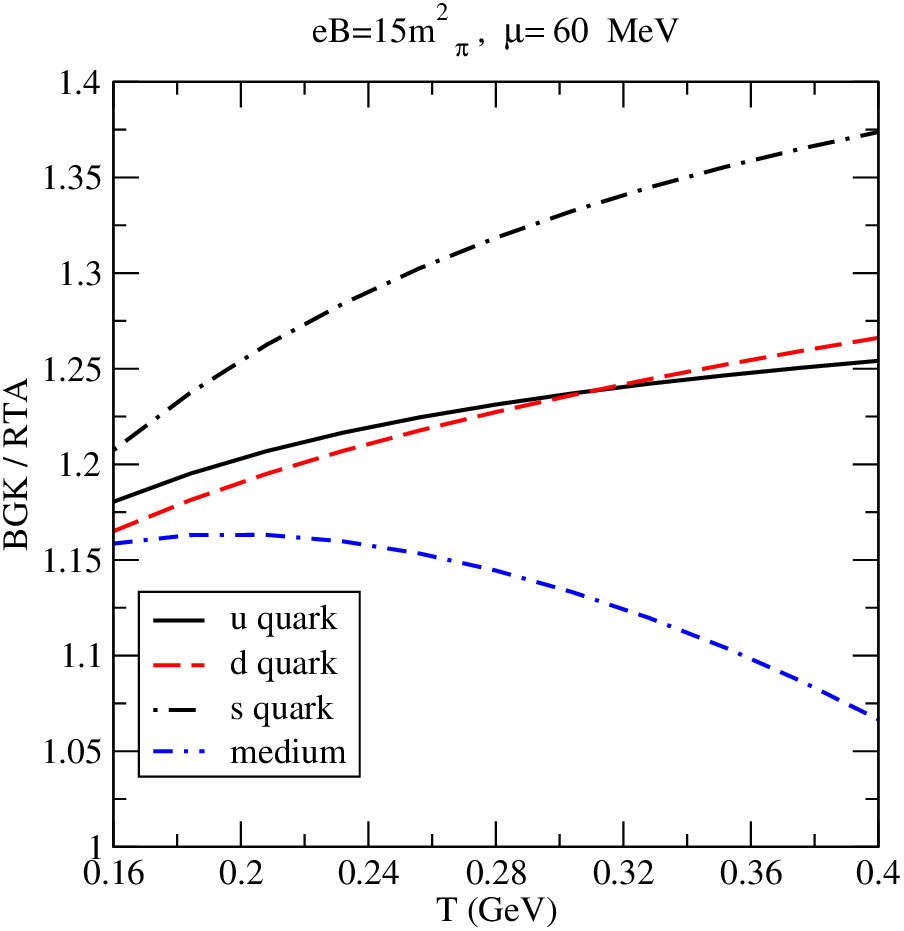}\\
(a) & \hspace{2.5cm}(b) 
\end{tabular}
\caption{{\bf Left panel:} Temperature dependence of
 Seebeck coefficient for the medium with BGK collision term
in strong  magnetic field for different strengths of
 $\mu$. {\bf Right panel:} Ratio of Seebeck 
 coefficient in BGK to that in RTA collision integral with temperature}   
\label{seebeck_B3} 
\end{figure}
\end{center}

 In Fig.~\ref{seebeck_weak1}, we investigate
how the BGK collision term modifies the
thermoelectric response  in the presence of
weak magnetic field ($eB=0.3~m_{\pi}^2$) for $u$ 
(left panel), $d$ (right panel) quarks. We notice that  
the magnitude of the Seebeck coefficient depends on 
 the chirality of the 
 quark quasiparticles. For the left-handed chiral modes,
 the ratio BGK/RTA is less than one for individual
 quark flavours 
 as well as for composite medium. The ratio
 is found in the range $0.97-0.98$ for the $u$ quarks in the temperature range $160<T<400$ MeV. In case of $d$ quarks and 
 medium, this ratio is around~$\sim 0.98$ [seen in Fig.~\ref{seebeck_weak3}
(a)]. This concludes that
 the BGK collision
 integral causes reduction in the Seebeck coefficients 
 in comparison to the RTA~\cite{Dey:arXiv'2204.06195}.
 On the other hand, BGK to RTA ratio is very 
 close to unity in case of right-handed modes, which 
 menifests that both the collision terms produce almost
 similar results
 [seen in Fig.~\ref{seebeck_weak3}
(b)]. 
 We have 
also studied the effects of the baryon asymmetry
on the thermoelectric phenomenon in Fig.~\ref{seebeck_weak2} (b)
for $\mu=40, 60$ and $80$ MeV and have noticed an 
 increase in Seebeck coefficient with $\mu$
 for both $L$ and $R$ modes. 
 
 Since BGK collision integral 
shows an improvement over RTA, it  gives more 
realistic  estimates of the 
transport coefficents {\em like} electrical
conductivity ($\sigma_{el}$), 
thermal conductivity ($\kappa$), 
shear ($\eta$) and bulk ($\zeta$) viscosities
 as compared to RTA~\cite{Khan:PRD104'2021,Khan:PRD106'2022}. 
The nonzero value of the Seebeck coefficient
modifies the electric current ( $J=\sigma_{el}E
-\sigma_{el}S\nabla T$) and thermal conductivity 
($\kappa=
\kappa_0-T\sigma_{el}S^2$) of the medium. Therefore, 
the BGK collision term will indirectly influence the charge and heat transport in the medium.
 Electrical conductivity plays an important role in the 
time evolution of the electromagnetic fields produced
in the noncentral collisions. Hence, the estimation 
of the electrical conductivity with realistic collision
integrals is of paramount importance
to understand the strength and life-span of the 
magnetic field during
the various stages of its evolution in the medium. 
The  magnetic field influences the particle production, 
dynamics of the heavy quarks and their bound states
(quarkonium) and
  many aspects of the QCD phase diagram~\cite{Kirill:AHEP2013,
Andersen:RMP88}. Similarly,  a more accurate 
understanding of the thermal conductivity is 
necessary to study the dynamics of the first order phase
transition~\cite{Skokov:NPA847'2010} and the chiral
critical point in the heavy ion collisions~
\cite{Joseph:PRC86'2012}.
It also govern the attenuation
of the sound in the medium via the Prandtl number.
In addition to charge and heat transport, 
 momentum transport   also
 gets affected  by 
the BGK collision integral so the hydrodynamic evolution of the 
medium may get influenced as shear and bulk viscosities act as input to the dissipative hydrodynamical equations. In
principle BGK collision term can affect the phenomenology
of the heavy ion collision in many ways.  

\vspace{8mm}
\begin{center}
\begin{figure}[H]
\begin{tabular}{cc}
\includegraphics[width=6cm]{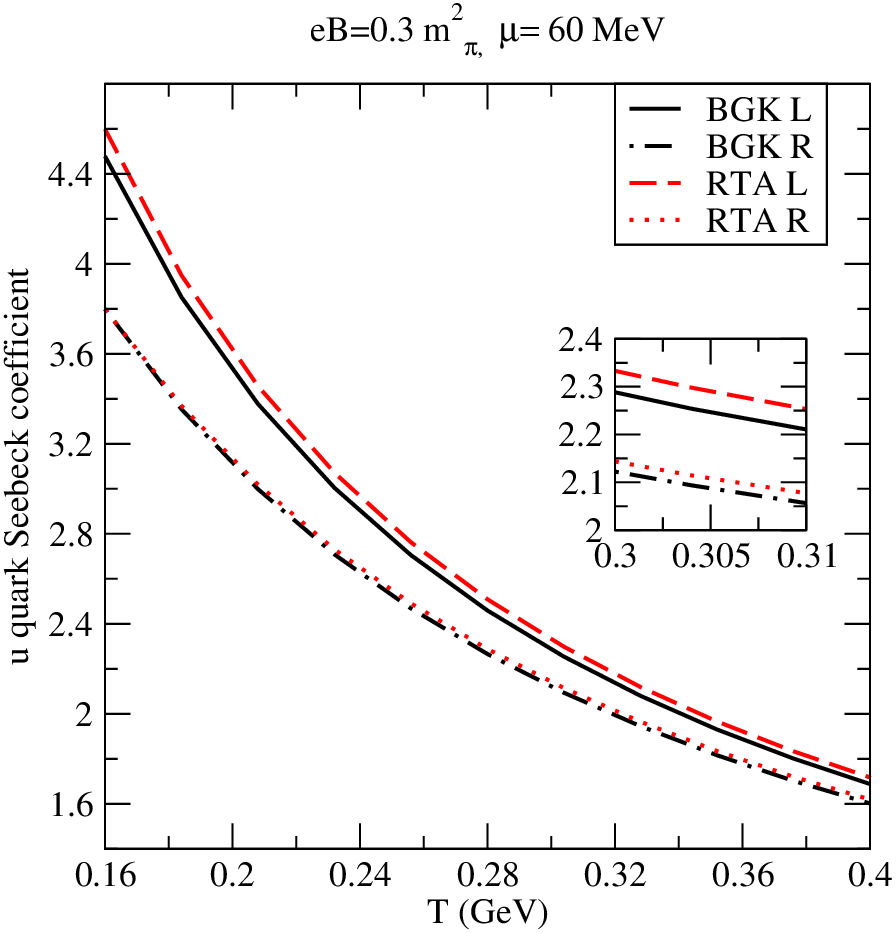}&
\hspace{2.5cm}
\includegraphics[width=6cm]{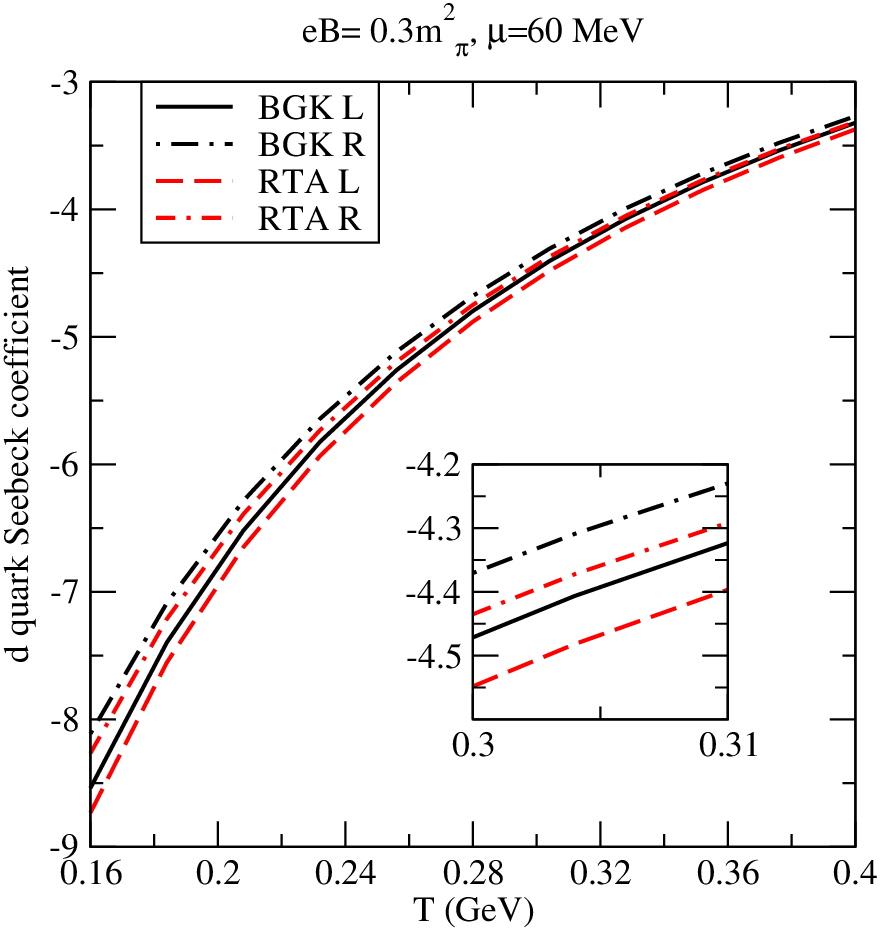}\\
(a) & \hspace{2.5cm}(b)
\end{tabular}
\caption{Variation of the Seebeck  coefficient with $T$ 
in the weak magnetic field  {\bf (a)} for u quarks, { \bf (b)} for d quarks.}
\label{seebeck_weak1} 
\end{figure}
\end{center}

\begin{center}
\begin{figure}[H]
\begin{tabular}{cc}
\includegraphics[width=6cm]{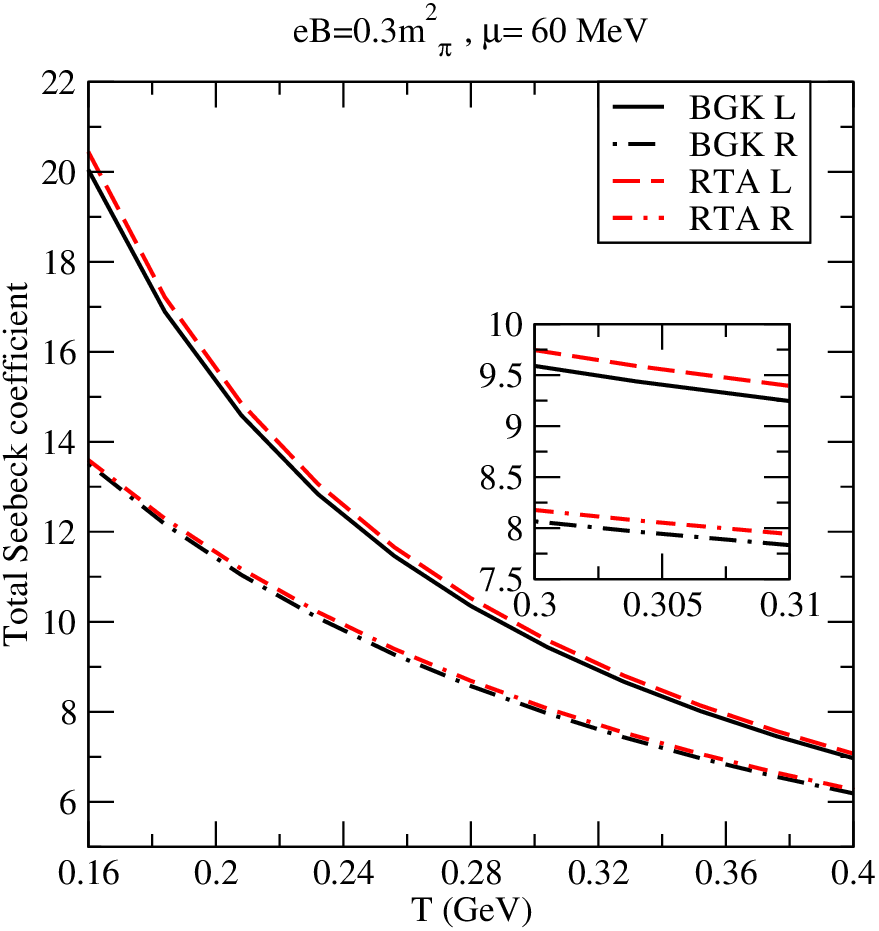}&
\hspace{2.5cm}
\includegraphics[width=6cm]{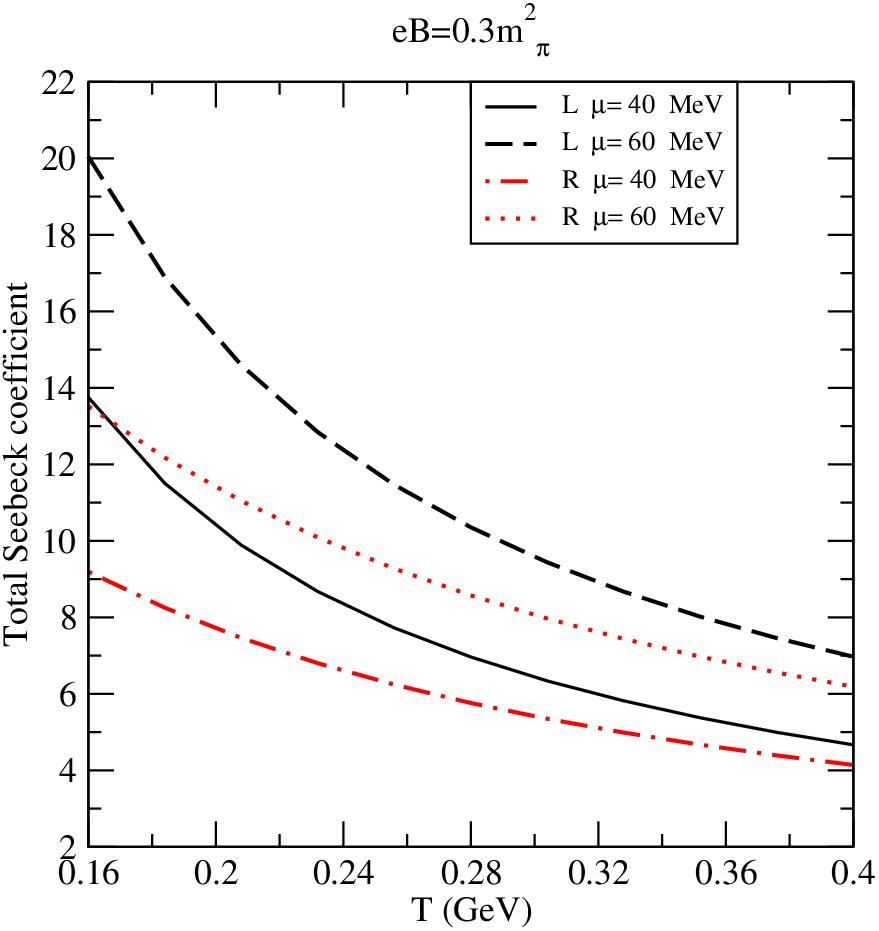}\\
(a) & \hspace{2.5cm}(b)
\end{tabular}
\caption{{\bf (a)} Variation of the Seebeck coefficient 
with $T$
for the composite medium,  {\bf (b)} Seebeck 
coefficient with respect to $T$ 
in the BGK collision term for different strengths of
 $\mu$. }
\label{seebeck_weak2} 
\end{figure}
\end{center}

\begin{center}
\begin{figure}[H]
\begin{tabular}{cc}
\includegraphics[width=6cm]{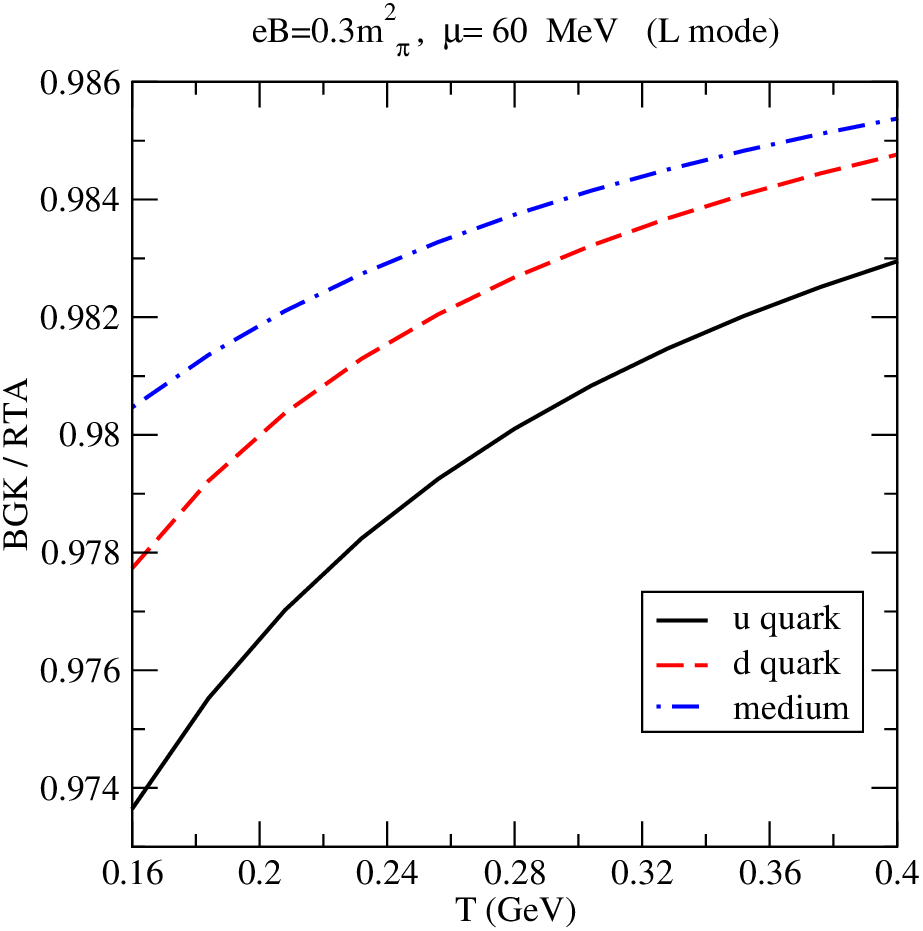}&
\hspace{2.5cm}
\includegraphics[width=6cm]{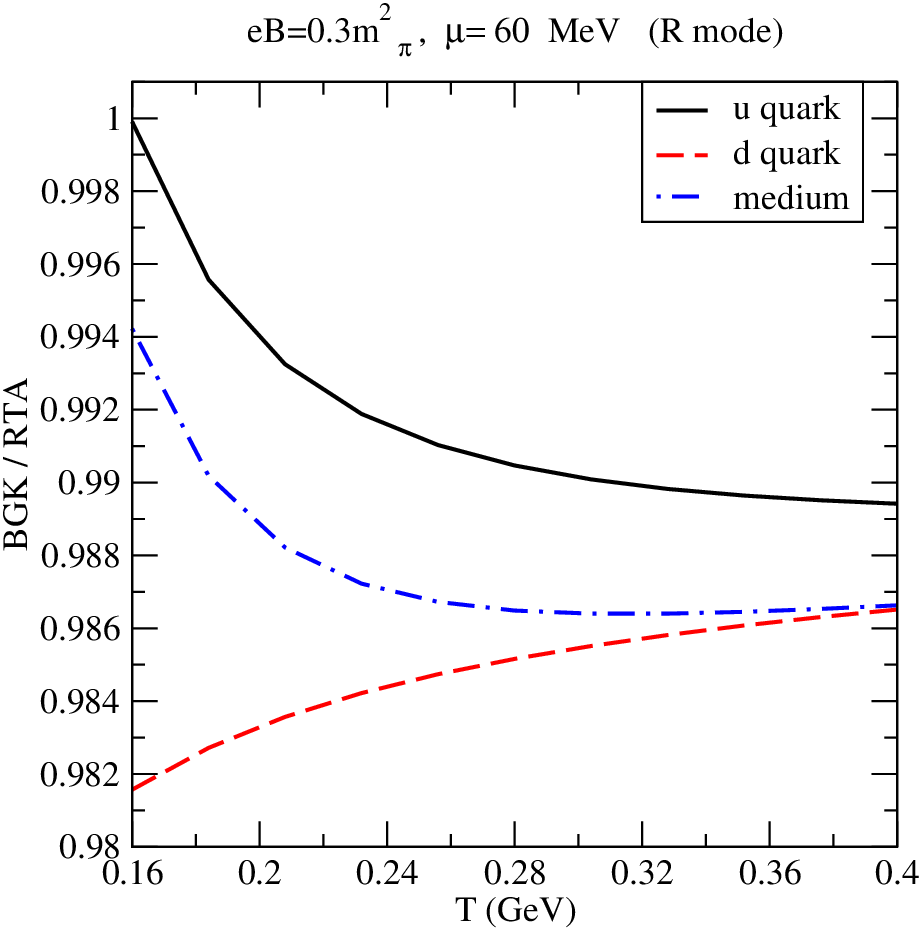}\\
(a) & \hspace{2.5cm}(b)
\end{tabular}
\caption{ { \bf (a)} Ratio of Seebeck coefficient in BGK to that in RTA 
 with $T$ for $L$ modes, 
 { \bf (b)} Ratio of Seebeck coefficient in BGK to that in RTA 
 with $T$ for $R$ modes.}
\label{seebeck_weak3} 
\end{figure}
\end{center}

\begin{center}
\begin{figure}[H]
\begin{tabular}{cc}
\includegraphics[width=6cm]{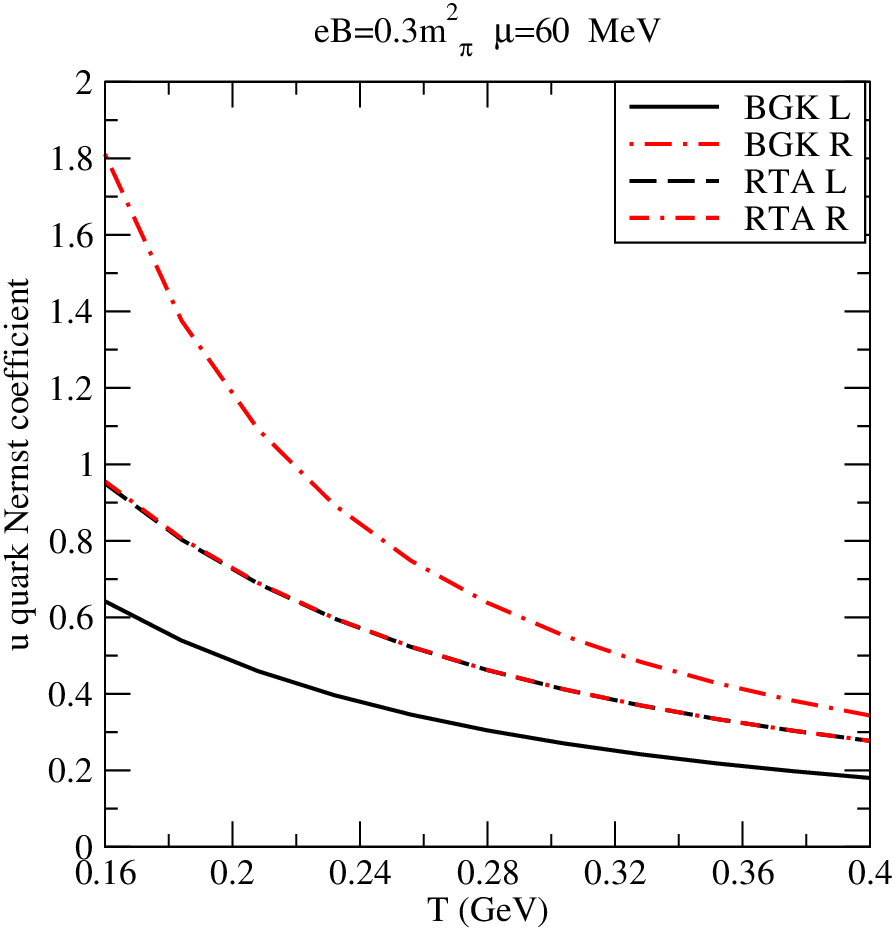}&
\hspace{2.5cm}
\includegraphics[width=6cm]{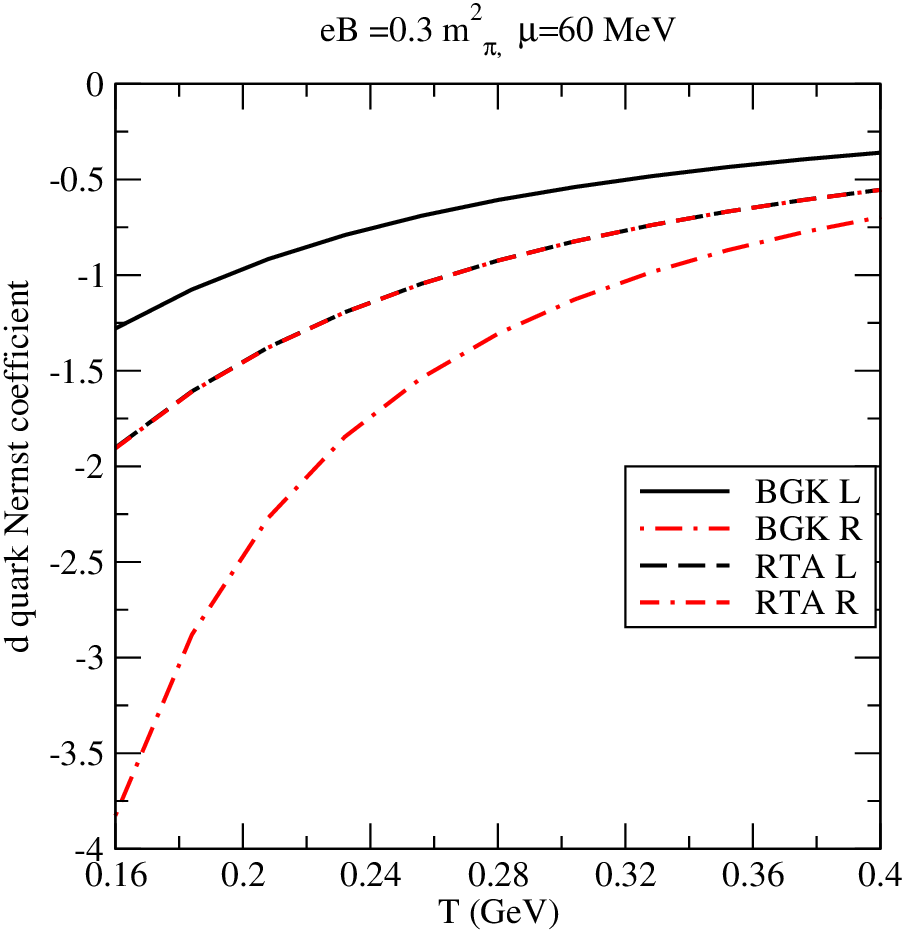}\\
(a) & \hspace{2.5cm}(b)
\end{tabular}
\caption{Variation of Nernst coefficient with respect to $T$ 
in the weak magnetic field {\bf (a)} for u quarks, { \bf (b)} for d quarks.}
\label{nernst_weak1} 
\end{figure}
\end{center}

\begin{center}
\begin{figure}[H]
\begin{tabular}{cc}
\includegraphics[width=6cm]{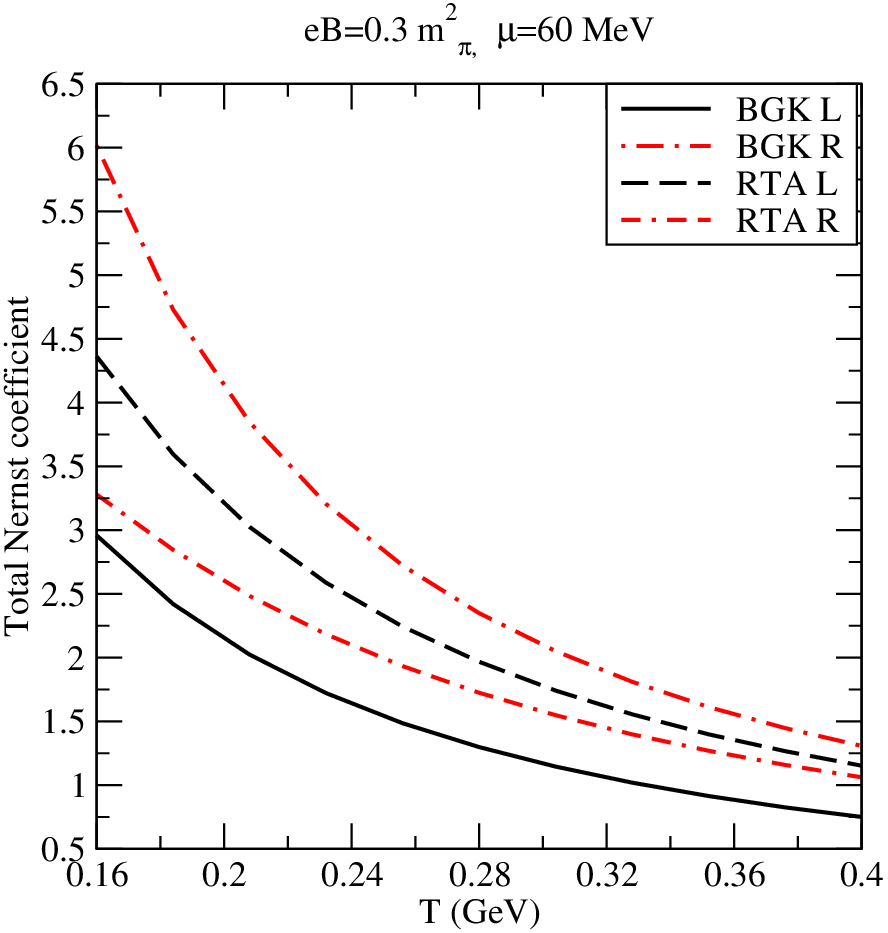}&
\hspace{2.5cm}
\includegraphics[width=6cm]{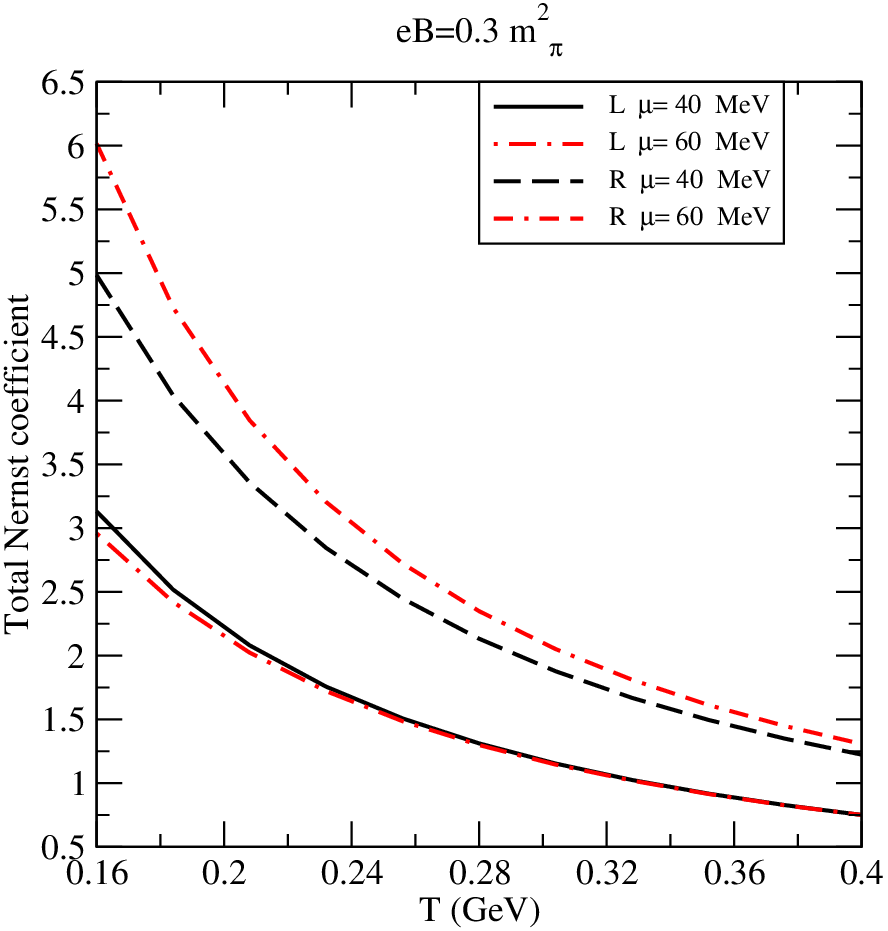}\\
(a) & \hspace{2.5cm}(b)
\end{tabular}
\caption{ {\bf (a)} Variation of Nernst coefficient with
respect to $T$
for the composite medium, 
{\bf (b)} Nernst coefficient with resect to $T$ 
 in the BGK collision term for different strengths of 
 $\mu$. }
\label{nernst_weak2} 
\end{figure}
\end{center}

\begin{center}
\begin{figure}[H]
\begin{tabular}{cc}
\includegraphics[width=6cm]{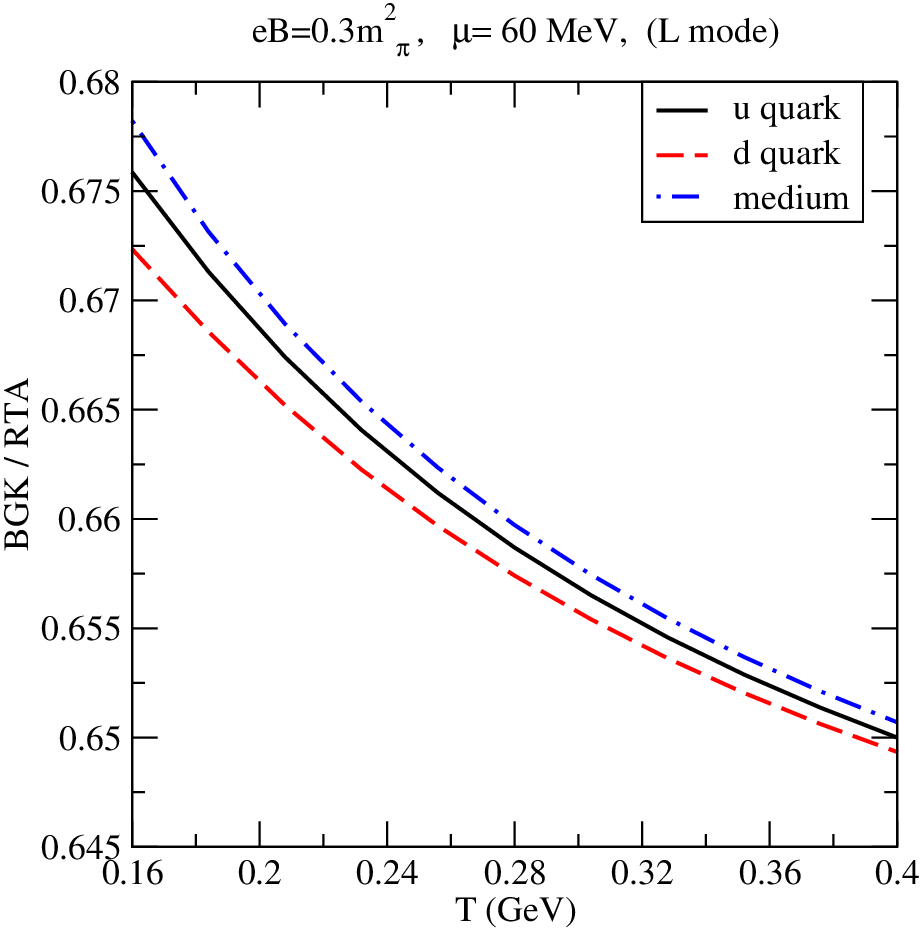}&
\hspace{2.5cm}
\includegraphics[width=6cm]{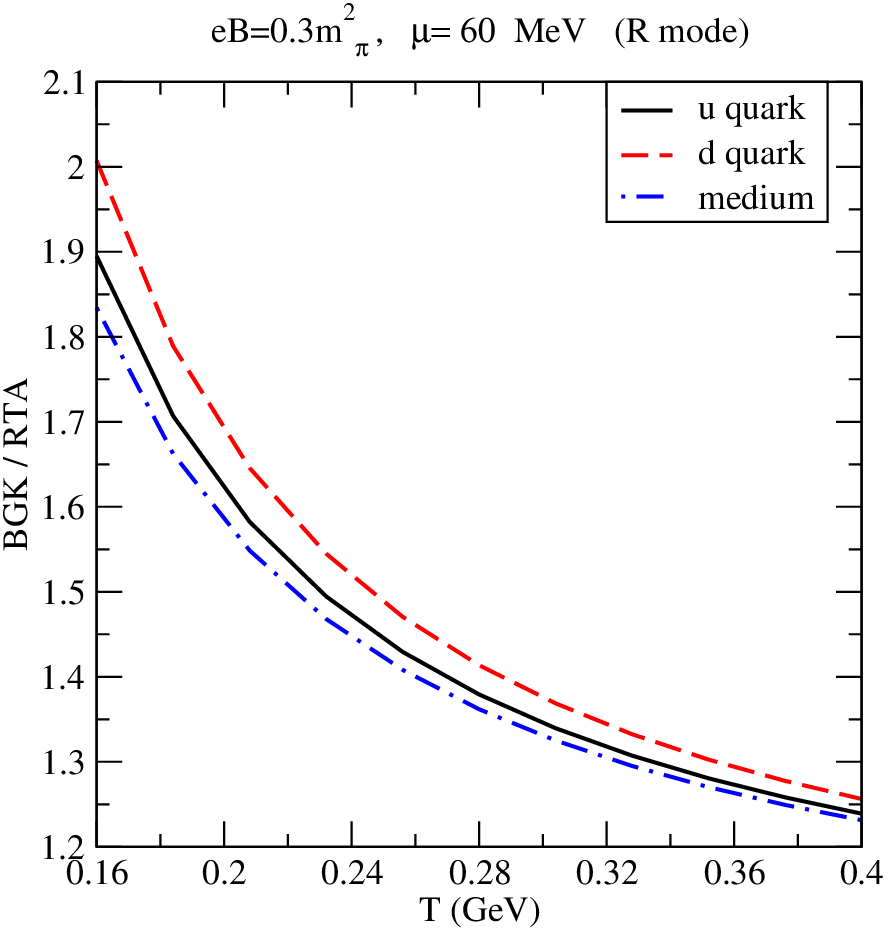}\\
(a) & \hspace{2.5cm}(b)
\end{tabular}
\caption{ { \bf (a)} Ratio of Nernst coefficient in BGK to that in RTA 
 with temperature for $L$ modes, 
 { \bf (b)} Ratio of Nernst coefficient in BGK to that in RTA 
 with temperature for $R$ modes.}
\label{nernst_weak3} 
\end{figure}
\end{center}
 In Fig.~\ref{nernst_weak1},
we have examined the
 collision integral
 dependence of the Nernst coefficient for $u$ 
 (left panel) and $d$ (right panel)       
 quarks. We observe that magnitude of 
 the Nernst coefficient
 gets changed drastically in the BGK collision term
 in comparison to RTA. The ratio BGK/RTA 
 for the Nernst coefficient is less than
 one in the case of left-handed modes, which shows 
 that BGK collision integral
 causes a reduction in the magnitude of 
 the Nernst coefficient~[see Fig.~\ref{nernst_weak3} (a)]. The
 ratio is in the range $0.67-0.65$ for the $u$ quarks, while 
 in the range $0.67-0.64$ and $0.68-0.65$
 for the $d$ quarks and for the medium, respectively. 
 In the case of $R$ modes,
 ratio becomes greater than unity and its value is found to be 
 in the range $1.89-1.23$ for $u$ quarks, $2.00-1.26$ for
 $d$ quarks and  $1.83-1.23$ for the 
 medium~[see Fig.~\ref{nernst_weak3} (b)]. 
One important observation we notice that 
 there is no difference
in the Nernst coefficient corresponding to 
 the $L$ and $R$ modes in RTA  for $u$ and $d$
 quarks but in the BGK the magnitude of 
 the Nernst coefficient is greater for $R$ modes. 
In Fig~\ref{nernst_weak2} (b), we study the 
$\mu$ dependance of the medium Nernst coefficient in BGK 
collision integral. It is not much visible in the case
of $L$ modes except near the transition temperature,
where it gets slightly reduced as $\mu$ increases. 
In the case of $R$ modes, the Nernst
coefficient increases with $\mu$ at a fixed value 
of temperature. \par

%%%%%%%%%%%%%%%%%%%%%%%%%%%%%%%%%%%%%%%%%%%%%%%%%%%%%%
\section{Conclusion}\label{five}
%%%%%%%%%%%%%%%%%%%%%%%%%%%%%%%%%%%%%%%%%%%%%%%%%%%%%%%%%
To conclude, we have investigated the thermoelectric 
response of a hot and magnetized  QCD medium produced 
in the non-central collisions at 
RHIC and LHC. We have employed 
Boltzmann transport equation 
linearized by the BGK collision integral,
which conserves the particle number and 
current instantaneously. We incorporate 
the medium effects  via dispersion 
relation wherein $T$, $\mu$ 
and $B$ dependent masses
have been calculated using the imaginary-time 
formalism of the finite temperature
  QCD. In the absence of $B$, the Seebeck 
coefficient  gets 
reduced in the BGK collision 
term for  $u$, $d$ and $s$ quarks whereas it gets 
 enhanced for the 
composite medium. In the strong $B$ background,
 the magnitude of the individual 
 as well as medium Seebeck coefficient
 get lifted in the BGK collision term. Seebeck coefficient
of the medium gets enhanced with the quark chemical 
potential in both the cases.   
 In addition to the Seebeck 
 coefficient, Nernst coefficient also appears in the 
 weak $B$. The  Seebeck 
 coefficient gets slightly reduced in the BGK 
 collision integral for $L$ modes, while
 for $R$ modes, both the collision integral
 give same results.
  On the other hand, Nernst coefficient gets changed 
  drastically in the BGK collision term and 
 its magnitude gets reduced (enhanced) for 
 $L$ ($R$) modes in comparison to RTA.
 Both Seebeck and Nernst coefficients
  increase with $\mu$ for both 
 $L$ and $R$ modes.\par
 
A nonvanishing 
Seebeck coefficient will modify the electric
as well as heat current in the medium. 
The electric current in
the presence of Seebeck effect becomes $J=\sigma_{el}E
-\sigma_{el}S\nabla T$, while thermal conductivity 
gets modified as  $\kappa=
\kappa_0-T\sigma_{el}S^2$. Both electrical and thermal 
conductivities should take positive values 
in accordance with the second law of thermodynamics.
{\em i.e. $T\partial_{\mu}S^{\mu}>0$}. Hence 
a positive Seebeck coefficient  
will always reduce the electric current and the 
thermal conductivity. It will be also interesting to take the 
thermoelectric effects into the account in
 the calculation of the entropy 
production, which has been completely 
neglected in~\cite{Gavin:NPA435'1985,Huang:PRD81'2010}. 
Moreover,
thermoelectric coefficients could 
also be relevant in the
context of the spin Hall effect (SHE). 
In SHE, a transverse spin current is generated 
due to the external electric field but the life-time 
of  such electric field produced in the 
heavy ion collisions could be too small to 
observe the SHE. The electric field produced 
due to the temperature gradients in the medium
may induce spin Hall effect  in a hot and dense
strongly interacting matter produced in
heavy-ion collisions~\cite{Liu:PRD104'2021}. 
So the study of the various implications of
thermoelectric effects 
in the hot and dense medium needs further investigation.  
 
 {\bf}

\section*{Acknowledgements}
S. A. Khan would like to thank  Debarshi Dey and  Pushpa Panday 
for various discussions.
\appendix
\appendixpage
\addappheadtotoc
\begin{appendix}
\renewcommand{\theequation}{A.\arabic{equation}}
\section{Derivation of equation \eqref{appendix_A}}
 BGK collision term 
is given by  \eqref{rbte_temp} as
\ba
C[f_i]&=&-p^{\mu}u_{\mu}\nu_i \left(f_i-\frac{n_i}{
n_{\mathrm eq,i}} f_{\mathrm eq,i}
\right)\nonumber\\
&=&-p^{\mu}u_{\mu}\nu_i \left(f_i-\frac{g_i 
\int_p (f_{\mathrm eq,i}+\delta f_i)}{n_{\mathrm eq}}
f_{\mathrm eq}\right)\nonumber\\
&=&-p^{\mu}u_{\mu}\nu_i \left(f-\frac{(g_i \int_p f_{\mathrm eq,i}
+g_i\int_p\delta f_i)}{n_{\mathrm eq,i}}f_{\mathrm eq,i}\right)\nonumber\\
&=&-p^{\mu}u_{\mu}\nu_i\left(\delta f_i-g_i n_{\mathrm eq,i}^{-1}
f_{\mathrm eq,i}\int_p\delta f_i\right).
\ea
\renewcommand{\theequation}{B.\arabic{equation}}
\section{Boltzmann Equation in the weak magnetic field}
The RBTE~\eqref{rbte} can be written with the BGK 
collision integral as  
\begin{eqnarray}
p^{\mu}\frac{\partial f_i}{\partial x^{\mu}}+q_i~F'^{\sigma}
\frac{\partial f_i}{\partial p^{\sigma}}= 
-p^{\mu}u_{\mu}\nu_i \left(f_i-\frac{n_i}
{n_{\mathrm eq,i}} f_{\mathrm eq,i}
\right)
\label{rbte_weak}
\end{eqnarray}
where $F'^{\sigma}=qF^{\sigma \rho}p_{\rho}=
(p^0 \vec{v}.\vec{F},p^0\vec{F})$, is the covariant form
 of the Lorenz force 
$\vec{F}=q_i(\vec{E}+\vec{v}\times \vec{B})$. We can write
Eqn.~\eqref{rbte_weak} using $F^{0i}=-E^{i}$ and 
$2F_{ij}=\epsilon_{ijk}B^k$ ($\epsilon_{ijk}$ is 
 anti-symmetric Levi-Civita tensor) as
\begin{eqnarray}
\frac{\partial f_i}{\partial t}+\vec{v}.
\frac{\partial f_i}{\partial \vec{r}}+
\frac{\vec{F}.\vec{p}}{p^0}
\frac{\partial f_i}{\partial p^0}+\vec{F}.
\frac{\partial f_i}{\partial \vec{p}}=
-\nu_i \left(f_i-\frac{n_i}
{n_{\mathrm eq,i}} f_{\mathrm eq,i}
\right)
\label{B_1}
\end{eqnarray}
considering $p^0$ as an independent variable
\begin{eqnarray}
\frac{\partial}{\partial \vec{p}} \rightarrow
\frac{\partial p^0}{\partial \vec{p}}\frac{\partial}{\partial p^0}+
\frac{\partial}{\partial \vec{p}}=
\frac{\vec{p}}{p^0}\frac{\partial}{\partial p^0}+
\frac{\partial}{\partial \vec{p}}
\end{eqnarray} 
Eqn.~\eqref{B_1} takes the form
\begin{eqnarray}
\vec{v}.
\frac{\partial f_i}{\partial \vec{r}}+
\vec{F}.\frac{\partial f_i}{\partial \vec{p}}=
-\nu_i \left(f_i-\frac{n_i}
{n_{\mathrm eq,i}} f_{\mathrm eq,i}
\right)
\end{eqnarray}

\renewcommand{\theequation}{B.\arabic{equation}}
\section{Seebeck coefficient in relaxation time 
approximation}\label{RTA_exp}
The Seebeck coefficient 
in the RTA collision term in $B=0$ case has been 
calculated as
~\cite{Dey:PRD102'2020}\\
\ba
S= \frac{1}{2 Tq}\left(\frac{L_1}{L_2}\right)
\ea 
\begin{eqnarray}
L_1&=&\int \frac{d^3p}{(2\pi)^3}~ 
\frac{p^2}{3\omega^2}\bigg \{(\omega -\mu) 
f_{\mathrm eq}\left( 
1-f_{\mathrm eq} \right)+
(\omega +\mu) 
\bar{f}_{\mathrm eq}\left( 
1-\bar{f}_{\mathrm eq} \right)\bigg\}\\
L_2&=&\int \frac{d^3p}{(2\pi)^3}~ 
\frac{p^2}{3\omega^2} \bigg \{
f_{\mathrm eq}\left( 
1-f_{\mathrm eq} \right)+\bar{f}_{\mathrm eq}\left( 
1-\bar{f}_{\mathrm eq} \right)\bigg \}
\end{eqnarray}
and for the case  of strong $B$ 
 as~\cite{Dey:PRD102'2020}
\ba
S_B=\frac{1}{2qT}\left(\frac{H_1}{H_2}\right)  
\ea
where
\begin{eqnarray}
H_1&=&\int {dp_3} \frac{p_3^2}{\omega^2}\tau^B
 \bigg \{(\omega-\mu)f_{\rm eq}^{B}\left( 1-f_{\rm eq}^{B}
  \right)+ (\omega+\mu)\bar{f}_{\rm eq}^{B}
  \left( 1-\bar{f}_{\rm eq}^{B} \right)
\bigg \}\\
H_2&=&\int {dp_3} \frac{p_3^2}{\omega^2}\tau^B
\bigg \{f_{\rm eq}^{B}\left( 1-f_{\rm eq}^{B} \right)
+ \bar{f}_{\rm eq}^{B}\left( 1-\bar{f}_{\rm eq}^{B} \right)
 \bigg \}
\end{eqnarray}
In other work~\cite{Dey:PRD104'2021}, 
the thermoelectric response of the
hot QCD medium has been studied in weak magnetic field, where 
 Seebeck  and Nernst coefficients are found to be
\ba
S&=& -\frac{(C_1C_3+C_2C_4)}{C_1^2+C_2^2},\\
N|B|&=&\frac{(C_2C_3-C_1C_4)}{C_1^2+C_2^2},
\ea
provided $C_1=qI_1$, $C_2=qI_2$, $C_3=\beta I_3$ and $C_4=\beta I_4$.
The integrals $I_1$, $I_2$,$I_3$ and $I_4$ are given by
\begin{eqnarray}
I_{1}&=&\int \frac{d^3p}{(2\pi)^3}~\frac{p^2}{3\epsilon^2}
\frac{ \tau}{(1+\omega_c^2 \tau^2)} \bigg \{
f_0(1-f_0)+\bar{f}_0(1-\bar{f}_0) \bigg \}\\
I_{2}&=&\int \frac{d^3p}{(2\pi)^3}~\frac{p^2}{3\epsilon^2}
\frac{\omega_c \tau^2}{(1+\omega_c^2 \tau^2)}\bigg \{
f_0(1-f_0)-\bar{f}_0(1-\bar{f}_0)\bigg \}\\
I_{3}&=&-\int \frac{d^3p}{(2\pi)^3}~\frac{p^2}{3\epsilon^2}
\frac{\tau}{(1+\omega_c^2 \tau^2)} \bigg \{
(\epsilon -\mu)f_0(1-f_0)-(\epsilon +\mu)\bar{f}_0(1-\bar{f}_0)\bigg \}\\
I_{4}&=&-\int \frac{d^3p}{(2\pi)^3}~\frac{p^2}{3\epsilon^2}
\frac{\omega_c \tau^2}{(1+\omega_c^2 \tau^2)}\bigg \{(\epsilon -\mu)
f_0(1-f_0)+(\epsilon +\mu)
\bar{f}_0(1-\bar{f}_0)\bigg \}
\end{eqnarray}

\end{appendix}

\end{document}